\documentclass[fleqn,usenatbib]{mnras}

\usepackage{newtxtext,newtxmath}

\usepackage[T1]{fontenc}

\DeclareRobustCommand{\VAN}[3]{#2}
\let\VANthebibliography\thebibliography
\def\thebibliography{\DeclareRobustCommand{\VAN}[3]{##3}\VANthebibliography}

\usepackage{graphicx}	
\usepackage{amsmath}	
\usepackage{ulem}
\usepackage{multirow}

\newcommand{\V}{\mathcal{V}}
\newcommand{\cl}{$C_{\ell}(\Delta\nu)$}
\newcommand{\maps}{$C_{\ell}(\nu_a,\nu_b)$}
\newcommand{\HI}{\ion{H}{i}}
\newcommand{\omb}{$[\Omega_{\HI} b_{\HI}]$}
\newcommand{\ombsq}{$[\Omega_{\HI} b_{\HI}]^2$}
\newcommand{\ellb}{\boldsymbol{\ell}}
\newcommand{\clb}{$C_{\ellb}(\Delta\nu)$}
\newcommand{\K}{\mathbf{K}}
\newcommand{\kfg}{\mathbf{k}_{\rm FG}}
\newcommand{\dv}{\mathbf{d}}

\defcitealias{Ch21}{Ch21} 
\defcitealias{P22}{Paper~I}
\defcitealias{AE23}{Paper~II}

\title[Foreground removal]{Towards $21$-cm intensity mapping at $z=2.28$ with uGMRT using the tapered gridded estimator III: Foreground removal}

\author[A. Elahi et al.]{Kh. Md. Asif Elahi,$^{1}$\thanks{E-mail:asifelahi999@gmail.com}
Somnath Bharadwaj,$^{1}$\thanks{E-mail:somnath@phy.iitkgp.ac.in}
Srijita Pal,$^{1}$
Abhik Ghosh,$^{2}$
Sk. Saiyad Ali,$^{3}$
\newauthor
Samir Choudhuri,$^{4}$
Arnab Chakraborty,$^{5}$
Abhirup Datta,$^{6}$
Nirupam Roy,$^{7}$
Madhurima Choudhury,$^{8}$
\newauthor
Prasun Dutta$^{9}$
\\
\\
$^{1}$ Department of Physics and Centre for Theoretical Studies, IIT Kharagpur, Kharagpur 721 302, India\\
$^{2}$ Department of Physics, Banwarilal Bhalotia College, Asansol, West Bengal-713303, India\\
$^{3}$ Department of Physics, Jadavpur University, Kolkata 700032, India\\
$^{4}$ Centre for Strings, Gravitation and Cosmology, Department of Physics, Indian Institute of Technology Madras, Chennai 600036, India\\
$^{5}$ Department of Physics and McGill Space Institute, McGill University, Montreal, QC, Canada H3A 2T8\\
$^{6}$ Discipline of Astronomy, Astrophysics and Space Engineering, Indian Institute of Technology Indore, Indore 453552, India\\
$^{7}$ Department of Physics, Indian Institute of Science, Bangalore 560012, India\\
$^{8}$ ARCO (Astrophysics Research Center), Department of Natural Sciences, The Open University of Israel, 1 University Road, PO Box 808, Ra’anana 4353701, Israel\\
$^{9}$ Department of Physics, IIT (BHU), Varanasi, 221005 India
}

\date{Accepted XXX. Received YYY; in original form ZZZ}

\pubyear{2023}

\begin{document}
\label{firstpage}
\pagerange{\pageref{firstpage}--\pageref{lastpage}}
\maketitle

\begin{abstract}
    Neutral hydrogen (\ion{H}{i}) $21$-cm intensity mapping (IM) is a promising probe of the large-scale structures in the Universe. However, a few orders of magnitude brighter foregrounds obscure the IM signal. Here we use the Tapered Gridded Estimator (TGE) to estimate the multi-frequency angular power spectrum (MAPS) $C_{\ell}(\Delta\nu)$ from a $24.4\,\rm{MHz}$ bandwidth uGMRT Band $3$ data at $432.8\,\rm{MHz}$. In $C_{\ell}(\Delta\nu)$ foregrounds remain correlated across the entire $\Delta\nu$ range, whereas the $21$-cm signal is localized within $\Delta\nu\le[\Delta \nu]$ (typically $0.5-1\,\rm{MHz}$). Assuming the range $\Delta\nu>[\Delta \nu]$ to have minimal $21$-cm signal, we use $C_{\ell}(\Delta\nu)$ in this range to model the foregrounds. This foreground model is extrapolated to  $\Delta\nu\leq[\Delta \nu]$, and subtracted from the measured $C_{\ell}(\Delta\nu)$. The residual $[C_{\ell}(\Delta\nu)]_{\rm res}$ in the range $\Delta\nu\le[\Delta\nu]$ is used to constrain the $21$-cm signal, compensating for the signal loss from foreground subtraction. $[C_{\ell}(\Delta\nu)]_{\rm{res}}$ is found to be noise-dominated without any trace of foregrounds. Using  $[C_{\ell}(\Delta\nu)]_{\rm res}$ we constrain the $21$-cm brightness temperature fluctuations $\Delta^2(k)$, and  obtain the $2\sigma$ upper limit $\Delta_{\rm UL}^2(k)\leq(18.07)^2\,\rm{mK^2}$ at $k=0.247\,\rm{Mpc}^{-1}$. We further obtain the $2\sigma$ upper limit $ [\Omega_{\ion{H}{i}}b_{\ion{H}{i}}]_{\rm UL}\leq0.022$ where $\Omega_{\ion{H}{i}}$ and  $b_{\ion{H}{i}}$ are the comoving \ion{H}{i}  density and bias parameters respectively. Although the upper limit is nearly $10$ times larger than the expected $21$-cm signal, it is $3$ times tighter over previous works using foreground avoidance on the same data.
\end{abstract}




\section{Introduction}

The bulk  of the neutral hydrogen (\HI{}) in the post Epoch of Reionization (post-EoR) era resides in dense clumps which are seen as Damped Ly-$\alpha$ Systems (DLAs) in quasar absorption spectra \citep{Lanzetta95, Wolfe95, Not, zafar, ho2021}. These dense \HI{} clumps are believed to be primarily associated with galaxies. The $21$-cm intensity mapping (IM) technique aims to measure the integrated redshifted $21$-cm emission from the \HI{} distribution instead of individually resolving the distant and faint galaxies \citep{BNS, BS01}. The IM signal is expected to trace the underlying matter distribution, authorising it as an excellent probe of the cosmological large-scale structures \citep{bh_sri2004, BA5, loeb08, bagla2010, ansari2012}. Apart from that, the $21$-cm IM signal can constrain the evolution of Dark Energy from the Baryon Acoustic Oscillation (BAO; \citealt{ Chang08, w08, battye2013, Bull15}) measurements, put independent limits to various cosmological parameters \citep{Visbal_2009, Bh09}, quantify non-Gaussianity \citep{Ali2006, Hazra2012}, and constrain the  Epoch of Reionization (EoR) models \citep{long2022}.

Appreciating the ample amount of cosmological possibilities, several ongoing experiments, such as the Giant Metrewave Radio Telescope (GMRT\footnote{\url{https://www.gmrt.ncra.tifr.res.in/}};  \citealt{swarup91}), MeerKAT\footnote{\url{https://www.sarao.ac.za/science/meerkat/}} \citep{Kennedy21, CunningtonMeerKLASS2023}, the Canadian Hydrogen Intensity Mapping Experiment (CHIME\footnote{\url{https://chime-experiment.ca/en/}}; \citealt{chimeIM}), the Ooty Wide Field Array (OWFA\footnote{\url{https://rac.ncra.tifr.res.in/ort.html}}; \citealt{OWFA}), the Hydrogen Intensity and Real-time Analysis eXperiments (HIRAX\footnote{\url{https://hirax.ukzn.ac.za/}}; \citealt{Newburgh16}), aim at measuring the post-EoR \HI{} IM signal. The forthcoming Square Kilometre Array (SKA-mid\footnote{\url{https://www.skatelescope.org/}}; \citealt{SKA15}) and the proposed Packed Ultrawide-band Mapping Array (PUMA; \citealt{puma}) further promise efficient and sensitive measurements of the \HI{} distribution across a wide redshift range. A few experiments (e.g. \citealt{chang10, SW13, Wolz2021, chime22, Cunnington23}) have successfully detected the IM signal in the low-redshifts $(z<1.3)$ by cross-correlating the \HI{} maps with galaxy surveys (such as, eBOSS; \citealp{Dawson_2016}). \cite{Paul23} have recently reported a direct detection of the IM signal at $z\approx0.32$ and $z\approx0.44$ using the MeerKAT interferometer. However, a high-redshift, auto-correlation detection of the faint $21$-cm IM signal is largely confronted by the $4-5$ orders of magnitude brighter Galactic and extragalactic foregrounds (see, e.g. \citealt{shaver99, ali, ghosh3}).

The foregrounds, which are believed to  originate from continuum sources, are expected to be spectrally smoother than the \HI{} $21$-cm signal which is a line emission (\citealt{BNS, dmat1}).  The Multifrequency Angular Power Spectrum (MAPS; \citealt{Zaldarriaga2004, santos05, KD07}) \maps{} jointly characterises the angular ($\ell$) and spectral ($\nu$) two-point statistics of the sky signal and promises the ability to  tell apart the foregrounds from the desired  IM signal \citep{liu12, Trott2022}. The MAPS \cl{}, with $\Delta\nu= \mid \nu_a - \nu_b \mid$,   is adequate to assess the IM signal when the frequency bandwidth under consideration is sufficiently small \citep{Mondal2018, Mondal19}. Considering  \cl{}, the foregrounds are expected to remain significantly correlated across the bandwidth, whereas the $21$-cm signal is expected to decorrelate rapidly with increasing $\Delta\nu$ \citep{Bharadwaj2003, BA5, KD07}. In principle, the foregrounds and the signal can be separated using these distinct decorrelation properties of MAPS \citep{ghosh1, ghosh2}. However, various factors like baseline migration \citep{Morales_2012},   ionospheric fluctuations and  calibration errors \citep{jais2020, jais22} introduce frequency-dependent structures. In particular, the point sources located at the periphery of the primary beam pattern  and at  the side lobes  show up as oscillations which are extremely difficult to model and remove from the \cl{} \citep{ghosh1}. Further, the period  of these oscillations  decreases  with increasing $\ell$ \citep{ghosh1} -- a feature that equivalently appears as the `foreground wedge'  in the cylindrical power spectrum (PS) $P(k_{\perp}, k_{\parallel})$ \citep{adatta10, Morales_2012, vedantham12, trott1, pober16}. The reader is referred to \citealt{P22}, hereafter \citetalias{P22}, for a discussion on how the oscillation in \cl{} and the wedge are related. Many  existing methods attempt to remove the foregrounds from the measured visibilities (e.g. \citealt{paciga11, adatta10, chapman12, mertens18,Trott2022}), assuming the smooth nature of the foregrounds. On the other hand, several works (e.g. \citealt{pober13, pober14, liu14a, liu14b, dillon14, dillon15, Pal20, Abdurashidova_2022})  have adopted the `foreground avoidance' strategy where only the region outside the foreground wedge is used to estimate the $21$-cm PS.

A novel approach to mitigate the wide-field foreground effects was first proposed by \cite{ghosh2}, who tapered the interferometer's sidelobe response to mitigate the oscillations in \cl{}. The Tapered Gridded Estimator (TGE) incorporates the tapering by convolving the visibilities with a suitably chosen window function \citep{samir14, samir16, samir17}. The TGE manifestly has three salient features. First, it tapers the sky response to mitigate the oscillations in \cl{}, thereby significantly reducing foreground contamination. Second, TGE uses gridded visibilities, reducing computational expenses and improving the signal-to-noise ratio (SNR). Third, the TGE  internally estimates the noise bias for an unbiased PS estimation. A two-dimensional (2D) version of TGE has been extensively used  to study the angular PS of the diffused Galactic foregrounds \citep{samir17a, samir20, Cha1, Mazumder20}  and also Galactic magnetohydrodynamic turbulence \citep{Preetha19, Preetha21}. Recently, a Tracking TGE \citep{Chatterjee2022} has been developed to analyse  drift scan interferometric observations. 

The observed visibility data typically has several frequency channels which are flagged to avoid Radio Frequency Interference (RFI) contamination. This introduces artefacts in the estimated PS, and several  state-of-the-art  algorithms \citep{Parsons_2009, Trott2016, Kern2021, Ewall-Wice2021, Kennedy2023} have been proposed to overcome this.  The MAPS-based TGE used for the present work naturally overcomes this problem by first estimating  \cl{} and using this to estimate $P(k_{\perp}, k_{\parallel})$.  This advantage arises because we  obtain  estimates of  $C_{\ell}(\Delta \nu)$ for every $\Delta \nu$ even in the presence of flagged frequency channels. This approach was proposed in \cite{Bh18} who validated it using simulations, and it was subsequently  demonstrated on observed  $150 \, {\rm MHz}$   GMRT data \citep{Pal20}.

The observations of the upgraded GMRT (uGMRT; \citealt{uGMRT}) Band 3 data considered  in this paper are described  in \cite{Cha2}, who also outline the initial analysis.  The present paper is the third in a series of papers which have used MAPS-based TGE to estimate the $21$-cm PS from this data. An earlier work, \cite{Ch21} (henceforward Ch21)  have used the one-dimensional  CLEAN algorithm \citep{Parsons_2009} to estimate the $21$-cm PS in delay space. In the first paper of the present series (\citetalias{P22}), we have  estimated the $21$-cm PS from a sub-band of this data combining the two available polarization (RR and LL).  In  \citealt{AE23} (hereafter \citetalias{AE23}) we  have estimated the $21$-cm PS by  cross-correlating  the two polarizations instead of combining them.  We find that this leads to  a substantial reduction in the level of foreground contamination  as compared to \citetalias{P22}.  Considering this estimated cross-correlation \cl{}, in the present work we introduce  a method to model  and remove the foregrounds. We show that this does away with the foreground wedge present in the earlier works, allowing us to use  the entire $(k_{\perp}, k_{\parallel})$ plane for estimating the $21$-cm PS. 

We closely follow the methodology of \cite{ghosh2} who attempted a foreground removal from the \cl{} estimated from a $610\,\rm{MHz}$ GMRT observation. Considering  the estimated  \cl{}, this is a combination of the $21$-cm signal, foregrounds and noise.  For the relevant $\ell$ range,  the $21$-cm signal  is presumed to be localized within $\Delta \nu \le 0.5 \, \rm{MHz}$, and at  large $\Delta \nu$ ($> 0.5 \, {\rm MHz})$ the signal's amplitude drops substantially \citep{BA5}. The key idea is that we may treat the large $\Delta \nu$ range as having very little (practically zero) $21$-cm signal, and use this to  model the foregrounds. This model is extrapolated to  small $\Delta \nu$ and  used to subtract out the foregrounds. The residual \cl{} are expected to contain only the  $21$-cm signal and noise. Be noted that we only use the small $\Delta \nu$ range ($\le  0.5 \, {\rm MHz}$) to put a constraint on the $21$-cm signal.

The rest of the paper is arranged as follows: Section~\ref{sec:data} briefly summarizes the observations, whereas Section~\ref{sec:FGR4maps} describes the TGE and the foreground removal strategies. The formalism for the cylindrical PS is presented in Section~\ref{sec:cylindricalps}, while the spherical PS and \omb{} are presented in Sections~\ref{sec:BMLE} and \ref{sec:omb} respectively. Our findings are summarized in Section~\ref{sec:conclusion}.

The cosmological parameters  $\Omega_{m} = 0.309$, $n_s = 0.965$, $h = 0.67$, $\Omega_{b}h^2 = 0.0224$, quoted in \cite{Planck18f}, have been used all through the paper.

\section{Data Description}
\label{sec:data}

We have used the uGMRT Band $3$ to carry out deep observations of the ELAIS-N1 field covering $1.8 \,\rm{deg}^2$ in a single pointing. The observations were spread over four nights during May $2017$, with a total observing time of $25$ hours including all calibration overheads. We have used an integration time of $2$ seconds and a baseband bandwidth of $200$ MHz $(300 - 500 \,{\rm MHz})$, divided into $8196$ frequency channels to achieve a spectral resolution ($\Delta \nu_c$) of $24.4\,\rm{kHz}$. 
\cite{Cha2} detailed the observation and an initial data processing, which is summarized here. We have used  the \textsc{aoflagger} \citep{Off10, Off12} for initial RFI flagging, followed by the \textsc{rflag} routine of the Common Astronomy Software Applications (\textsc{casa}; \citealt{casa07}) to flag residual low-level RFIs from calibrated data. Regarding the calibration, only direction-independent calibration is done on the data using \textsc{casa}, and no polarization calibration is performed. 
After flagging and calibration of the visibility data, unresolved and compact sources brighter than $100 \mu Jy$ are identified and subtracted out using the \textsc{casa} routine \textsc{uvsub}. The residual visibility data after \textsc{uvsub} is used for all the subsequent analyses reported in this work. 

We have split a $24.4$ MHz bandwidth subset from the above data at the central frequency $\nu_c = 432.84 \, \rm{MHz}$  which has  a visibility r.m.s. of $0.43 \, \rm{Jy}$ (\citetalias{P22}). We also note that the initial rounds of flagging and calibration lead to $\sim55\%$ of this data being flagged (\citetalias{P22}). We have analysed this data in \citetalias{P22} and \citetalias{AE23}, and we 
extend our study utilizing the same data here. 
Similar to \citetalias{AE23}, for our analysis we have used only the baselines with $\mathbf{U} \leq 1\, k\lambda$ where the `$uv$-coverage' is adequately dense and more or less uniform. 
Note that in \citetalias{P22} we have considered a broader baseline range $(\mathbf{U}\leq3\, k\lambda)$.
 
We have used the notation $\V{}_i^{x}(\nu_a)$ to refer to the visibility data corresponding to baseline $\mathbf{U}_i$, in polarization state $x$ (RR and LL circular polarization correlation products for the present data) and at an observing frequency of $\nu_a$.

\section{Foreground Removal from MAPS}
\label{sec:FGR4maps}

\subsection{The Cross TGE}
\label{sec:maps}

The multi-frequency angular power spectrum (MAPS) \maps{}  is formally defined as
\begin{equation}
    C_{\ell}(\nu_a, \nu_b) = \big\langle a_{\ell {\rm m}} (\nu_a)\, a^*_{\ell{\rm m}} (\nu_b) \big\rangle 
    \label{eq:cl}
\end{equation}
where $a_{\ell {\rm m}} (\nu)$ are the coefficients when we expand $\delta T_{\rm b} (\hat{\mathbf{n}},\,\nu)$ the brightness temperature fluctuations in terms of spherical harmonics $Y_{\ell}^{\rm m}(\hat{\mathbf{n}})$ on the sky. The ensemble average  $\langle  ... \rangle$ refers to different statistically independent realizations of the random  field  $\delta T_{\rm b} (\hat{\mathbf{n}},\,\nu)$. 

The `Cross'  Tapered Gridded Estimator (TGE) for MAPS which cross-correlates the calibrated visibility data $\V{}_i^{x}(\nu_a)$  measured in two orthogonal polarizations $(x=RR, LL)$ is described in \citetalias{AE23}, and we briefly summarize this here. We first calculate  $\V_{cg}^{x}(\nu_a)$ the convolved-gridded visibility for every grid point $\mathbf{U}_g$ on a rectangular grid in the $uv$-plane using 
\begin{equation}
    \V_{cg}^{x}(\nu_a) = \sum_i  \tilde{w}(\mathbf{U}_g-\mathbf{U}_i) \, \V{}_i^{x}(\nu_a) \, F_i^{x}(\nu_a) \,.
    \label{eq:vcgx}
\end{equation}
where  $F_i^{x}(\nu_a)$, which takes values $0$ or $1$,  accounts for the flagged channels, and $\tilde{w}(\mathbf{U})$ is the Fourier transform of the window function ${\mathcal W}(\theta)=e^{-\theta^{2}/[f \theta_{0}]^2}$  which is introduced  to taper the sky response away from the beam centre. Here $\theta_0$ is $0.6$ times the FWHM of the PB, and following \citetalias{P22} we have used $f=0.6$ for the present analysis. 

The Cross TGE is defined as 
\begin{align}
  \hat{E}_g (\nu_a,\,\nu_b) = M_g^{-1}(\nu_a,\nu_b) {\mathcal Re} \Big[  &  \V_{cg}^{RR}(\nu_a)  \V_{cg}^{*LL}(\nu_b)  \nonumber \\ 
    & + \V_{cg}^{LL}(\nu_a) \V_{cg}^{*RR}(\nu_b)   \Big]  \,. 
    \label{eq:crossmaps}
\end{align}
where $ M_g(\nu_a,\nu_b) $ is a normalization factor which we have estimated using simulations corresponding to $C_{\ell}(\nu_a,\nu_b)=1$  referred to as unit MAPS or uMAPS. The simulated  visibilities $ [\V_i^{x}(\nu_a)]_{\rm{uMAPS}}$ are used to estimate 
\begin{align}
    {M}_g(\nu_a,\nu_b) = {\mathcal Re} \Big[\V_{cg}^{RR} & (\nu_a)  \V_{cg}^{*LL}(\nu_b)  \nonumber \\ 
    & + \V_{cg}^{LL}(\nu_a) \V_{cg}^{*RR}(\nu_b)   \Big]_{\rm {uMAPS}} \,.  
    \label{eq:uMAPS}
\end{align}
We have used multiple realizations ($50$ in this work) of the uMAPS to decrease the statistical uncertainties in the estimated $M_g$.

The estimator (equation~\ref{eq:crossmaps}) is unbiased i.e.  $\langle \hat{E}_g\rangle = C_{\ellb_g}$  at each grid ${\bf U}_g$ where $\ellb_g= 2\,\mathrm{\pi}\, {\bf U}_g $. We subsequently use the notation $C_{\ellb}$ and $\ellb$ to denote $C_{\ellb_g}$ and $\ellb_g$ respectively.  The cosmological $21$-cm signal is  statistically  isotropic on the sky plane i.e.  $C_{\ellb} \equiv C_{\ell}$ where $\ell= \mid \ellb \mid$ corresponds to an angular multipole. It is then possible to do an annular binning of the estimated $\hat{E}_g$ in the $uv$-plane. We have taken this approach in \citetalias{P22} and \citetalias{AE23} to increase the SNR, and also to reduce the data volume for a PS estimation.  However, the isotropy does not hold for the foregrounds, and the foreground contribution differs  from grid point to grid point. We find it advantageous to individually model and remove the foregrounds separately at each grid point.  Hence we avoid binning $C_{\ellb}$ at this stage, and we perform the binning only after foreground removal.

\subsection{The predicted signal}
\label{sec:predicted}

Here we follow  \citetalias{AE23} and consider $C_{\ell}(\Delta \nu)$ instead of $C_{\ell}(\nu_a, \nu_b)$. This assumes  the statistics of the $21$-cm signal to be ergodic along the line-of-sight (LoS) which is quite justified  given the relatively small redshift interval of $\Delta z = 0.19$ centred around  $z = 2.28$.  We have used 
\begin{align}
    \left[C_{\ell_a}(\Delta\nu_n)\right]_T =  \left[\Omega_{\HI} b_{\HI}\right]^{2}  & \frac{\bar{T}^{2}}{\mathrm{\pi} r^2} \int_{0}^{\infty}   d k_{\parallel} \cos(k_{\parallel}r^{\prime}\Delta\nu_n)  \, \nonumber \\ 
    \times & \, \mathrm{sinc}^2(k_{\parallel}r^{\prime}\Delta\nu_c/2) \,     
    P_m(k_{\perp a}, k_{\parallel}) \, 
    \label{eq:cl_Pk_sinc}
\end{align}
to calculate the predicted    $\left[C_{\ell_a}(\Delta\nu_n)\right]_T$  corresponding to the $21$-cm signal. Here the equation~(\ref{eq:cl_Pk_sinc}) has been used in a slightly modified form in \citetalias{AE23} who took it from \cite{BA5}. In equation~(\ref{eq:cl_Pk_sinc}), $P_{m}(\mathbf{k_{\perp}}, k_{\parallel})$ is the dark matter PS in redshift space at the wave vector  $\mathbf{k}$ whose LoS and perpendicular components are $k_{\parallel}$ and $\mathbf{k_{\perp}} = {\ellb}/{r}$ respectively, and the $\rm{sinc}$ function takes care of the finite width $\Delta\nu_c$ of the frequency channels. Considering $z=2.28$, we find that the comoving distance $r$, its frequency derivative $r^{\prime}$ have the values  $5703\,{\rm Mpc}$, $9.85\,{\rm Mpc/MHz}$, respectively (\citetalias{AE23}).
The mean brightness temperature  \citep{BNS,BA5}
\begin{equation}
    \bar{T}(z) = 133 \,{\rm mK}\,(1+z)^{2}\,\bigg(\frac{h}{0.7}\bigg)\,\bigg(\frac{H_{0}}{H(z)}\bigg)
    \label{eq:tbar}
\end{equation}
is found to have a value of $406\,\rm{mK}$ at this redshift. $\Omega_{\HI}$ and $b_{\HI{}}$  here are the comoving \HI{} mass density in units of the present critical density and the bias parameter, respectively. We have used $[\Omega_{\HI} b_{\HI}]=10^{-3} $ which is supported by various observations (e.g. \citealt{Rhee2018} and references therein) and simulations (e.g. \citealt{Deb16}). For the matter PS $P_{m}(\mathbf{k})$ we have used the fitting formula presented in \cite{Eisenstein_1998} ignoring the effect of redshift space distortion (RSD; \citealt{BA5}). 

\begin{figure}
    \centering
    \includegraphics[width=\columnwidth]{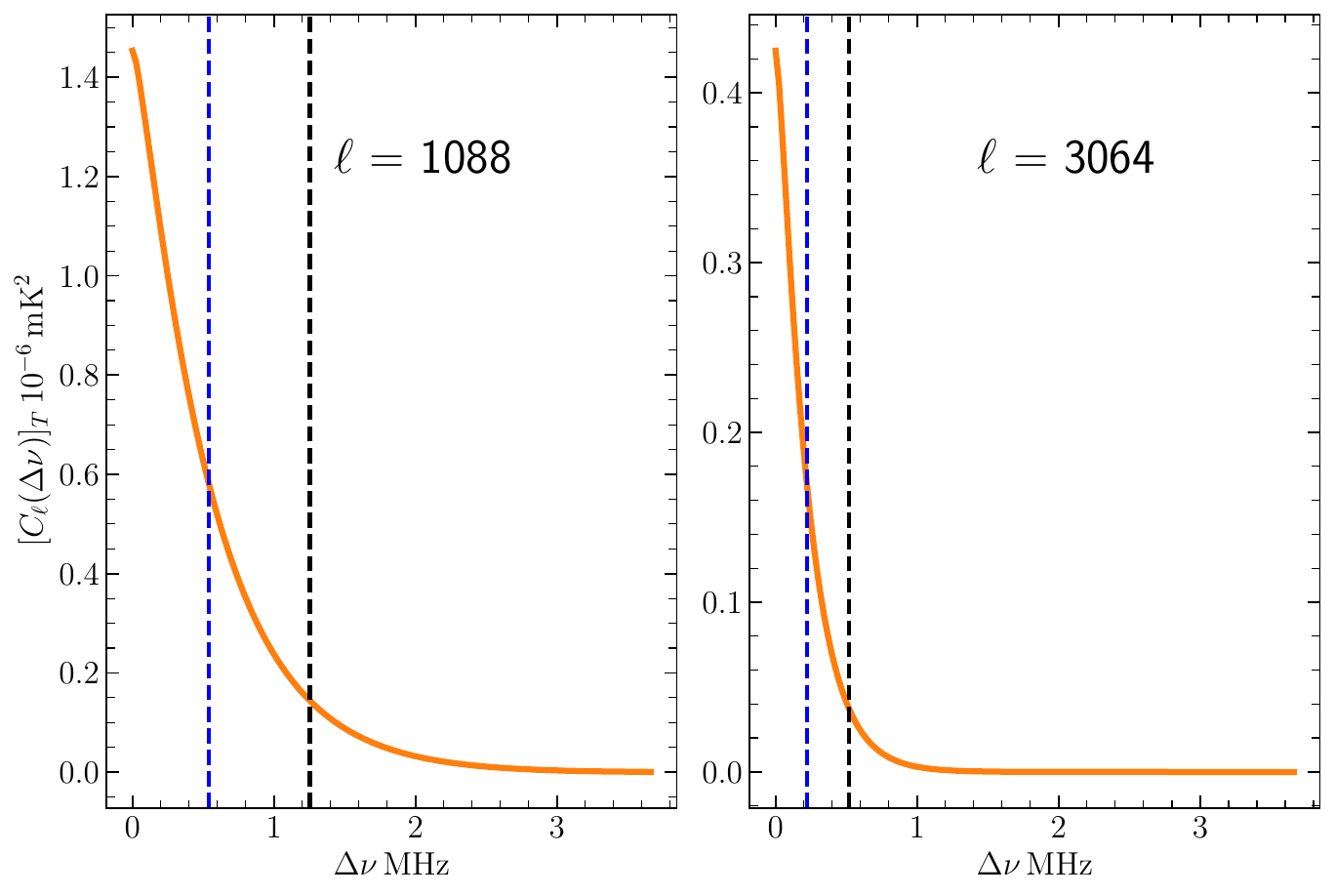}
    \caption{The theoretically predicted $21$-cm signal corresponding to $[\Omega_{\HI} b_{\HI}]=10^{-3} $ is shown. The blue and the black dashed vertical lines show $[\Delta\nu]_{0.4}$ and $[\Delta\nu]_{0.1}$ respectively for the corresponding $\ell$ values.}
    \label{fig:predicted_signal}
\end{figure}

Figure~\ref{fig:predicted_signal} shows $[C_{\ell}(\Delta\nu)]_T$ for two representative $\ell$ values. Considering both the panels, we see that $[C_{\ell}(\Delta\nu)]_T$ peaks at $\Delta\nu=0$, declines with increasing $\Delta \nu$ and  $[C_{\ell}(\Delta\nu)]_T \sim 0$ when $\Delta \nu \gtrsim 2\, {\rm MHz}$. We define $[\Delta\nu]_{0.4}$ and  $[\Delta\nu]_{0.1}$ respectively  as the values of $\Delta\nu$ where the amplitude of $[C_{\ell}(\Delta\nu)]_T$ falls  to $40\%$ and $10 \%$ of its peak value $[C_{\ell}(0)]_T$. In each panel, the blue and the black dashed vertical lines indicate $[\Delta\nu]_{0.4}$ and $[\Delta\nu]_{0.1}$ respectively for the corresponding $\ell$ value.  We see that the  $21$-cm signal is predominantly  localised within a small range of frequency separations,  and there is a very little $21$-cm signal at $\Delta \nu >  [\Delta\nu]_{0.1}$. The quantities   $[\Delta\nu]_{0.4}$ and  $[\Delta\nu]_{0.1}$  provide   estimates of this range of frequency separations within which the $21$-cm signal is localised.

The left panel of Figure~\ref{fig:predicted_signal} shows $[C_{\ell}(\Delta\nu)]_T$ for $\ell = 1088$. This has a peak value of $\sim1.4\times 10^{-6}\,\rm{mK}^2$ at $\Delta\nu=0$, and $[\Delta\nu]_{0.4} \approx 0.54 \, \rm{MHz} $ and $[\Delta\nu]_{0.1} \approx 1.25 \, \rm{MHz}$ respectively. The right panel shows $[C_{\ell}(\Delta\nu)]_T$ for a comparatively larger $\ell = 3064$ value. For this $\ell$, the peak value is $\sim0.4\times 10^{-6}\,\rm{mK}^2$, and we have  $[\Delta\nu]_{0.4} \approx 0.22 \, \rm{MHz}$ and $[\Delta\nu]_{0.1} \approx 0.52 \, \rm{MHz}$ respectively.  We see that  with increasing $\ell$, the amplitude of $[C_{\ell}(\Delta\nu)]_T$ goes down, and it also decorrelates faster {\it i.e.} $[\Delta\nu]_{0.4}$ and $[\Delta\nu]_{0.1}$ get smaller.

\subsection{Foreground modelling and removal: Polynomial fitting (PF)}
\label{sec:fgmodelling}

The foregrounds, which are believed to  originate from continuum sources, are expected to be spectrally smooth (\citealt{BNS, dmat1}). However,  baseline migration \citep{Morales_2012, vedantham12} and the frequency dependence of the  PB \citep{ghosh1} introduce additional chromaticity in the measured \clb{}. Regardless, the foregrounds are expected to be present over the entire  $\Delta\nu$-range, whereas the $21$-cm signal is largely localized within a small  $\Delta\nu $ range,  and there is very little $21$-cm signal for  $\Delta \nu > [\Delta \nu]_{0.1}$.  Here we assume that in the range $\Delta\nu > [\Delta\nu]_{0.1}$  the measured \clb{} can be  modelled as
\begin{equation}
    C_{\ellb}(\Delta\nu) = \left[C_{\ellb}(\Delta\nu)\right]_{\rm{FG}} + [\textrm{Noise}] 
    \label{eq:polyfitmodel} 
\end{equation}
where  $\left[C_{\ellb}(\Delta\nu)\right]_{\rm{FG}}$   and $[\textrm{Noise}]$  are the foreground and noise contributions respectively. In our first approach, we have used polynomial fitting (PF) to model the foregrounds and subtract these out. The estimated \clb{} is, by construction,  a symmetric function of $\Delta\nu$, i.e., $C_{\ellb}(\Delta\nu) = C_{\ellb}(-\Delta\nu)$. We have chosen an even polynomial in $\Delta \nu$ to model the foregrounds, 
\begin{equation}
    \left[C_{\ellb}(\Delta\nu)\right]_{\rm{FG}} = \sum_{m=0}^n a_{2m} \,\, (\Delta\nu)^{2m}
    \label{eq:FGmodel}
\end{equation}
where the coefficients $a_{2m}$ are the free parameters of our foreground model.  We have fitted equation~(\ref{eq:FGmodel}) to the measured  \clb{} in  the range $\Delta\nu > [\Delta\nu]_{0.1}$, and used this  to obtain the best-fit values of the parameters $a_{2m}$. The idea is to  extrapolate the best-fit $\left[C_{\ellb}(\Delta\nu)\right]_{\rm{FG}}$ to $\Delta\nu  \le  [\Delta\nu]_{0.1}$ where we use  it to subtract  out the foregrounds 
\begin{equation}
    \left[C_{\ellb}(\Delta\nu)\right]_{\rm{res}} = C_{\ellb}(\Delta\nu) - \left[C_{\ellb}(\Delta\nu)\right]_{\rm{FG}} \, .
    \label{eq:residual}
\end{equation}
Subsequent to foreground subtraction, the entire analysis is restricted to the range $\Delta\nu  \le  [\Delta\nu]_{0.1}$ which was not included in  PF.  We have used $ [C_{\ellb}(\Delta\nu)]_{\rm{res}}$ to measure (or constrain) the \HI{} $21$-cm signal. 

We have used maximum likelihood to find the best-fit parameters $a_{2m}$ and their error-covariance $C_a$. The error covariance $C_a$ also allows us to quantify the uncertainty in the best-fit $\left[C_{\ellb}(\Delta\nu)\right]_{\rm{FG}}$. We assume that the polynomial coefficients $a_{2m}$ follow a multivariate normal distribution (MND) with mean values as obtained from the maximum likelihood solutions and covariance $C_a$, i.e. $a_{2m} \sim \mathcal{N}(a_{2m}^{\rm{MLE}}\,,\, C_a)$. We generate $1000$ realizations of the polynomial coefficients from the MND and obtain the fits for each realization. The mean of the fits is by construction the best-fit foreground model $\left[ C_{\ellb}(\Delta\nu) \right]_{\rm{FG}}$, whereas the standard deviation across the realizations yields the error in the fits. These fitting errors are added to the error budget for each data point of the residuals $ [C_{\ellb}(\Delta\nu)]_{\rm{res}}$.

The parameter $n$,  which decides the order of the fitting polynomial, is an extra  free parameter in our foreground model. We have considered different values $0 \le n  \le 10$, and chosen the one that best matches the measured \clb{} in the range $\Delta\nu  \le  [\Delta\nu]_{0.1}$ and hence gives the smallest residual \clb{}. We have quantified this using the mean squared error (MSE)
\begin{equation}
    \rm{MSE} = \frac{\sum_{i}^{\rm{k}} \left[C_{\ellb}(\Delta\nu_i)\right]^2_{\rm{res}}} {(k-n)}
    \label{eq:mse}
\end{equation}
 where $\rm{k}$ is the number of $\Delta\nu$ data points in the range $\Delta\nu \le  [\Delta\nu]_{0.1}$. We have used the value of  $n$ which minimises MSE. 

\begin{figure}
    \centering
    \includegraphics[width=\columnwidth]{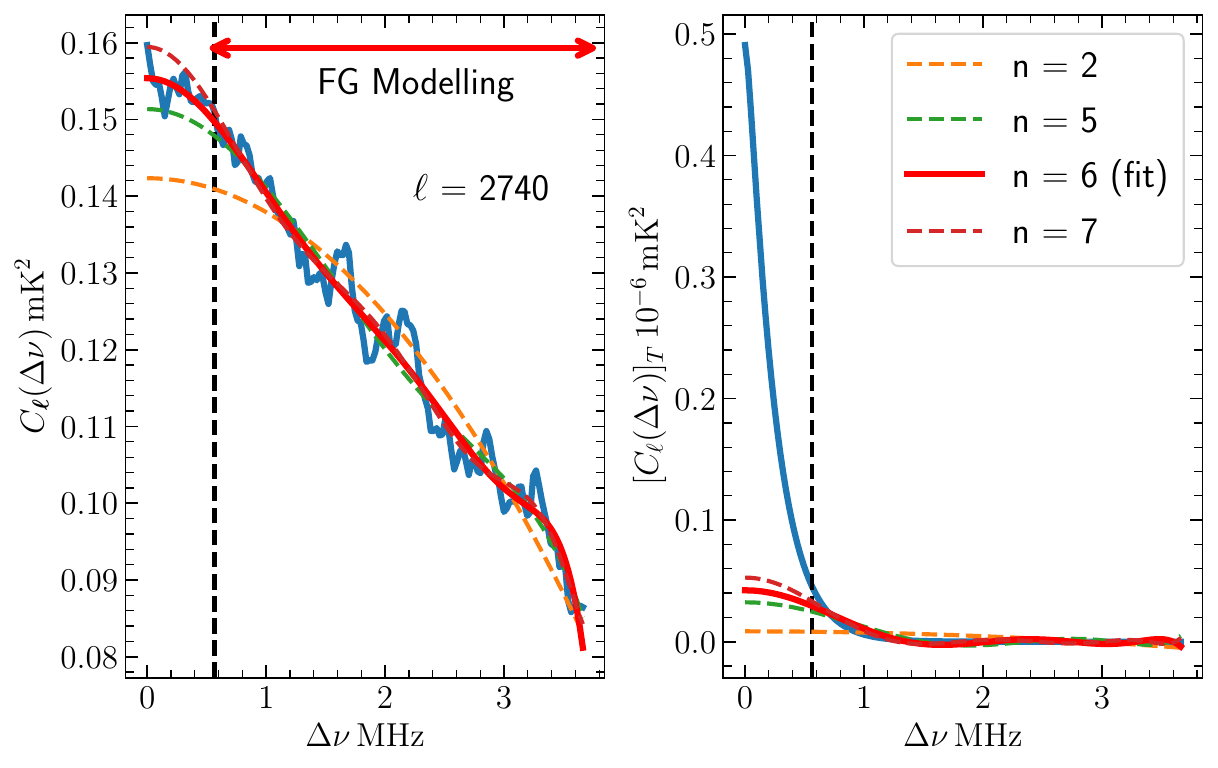}
    \caption{The left and the right panels respectively show (blue solid line) the measured \clb{} and $[C_{\ell}(\Delta\nu)]_T$ for a representative $\ellb = 2740$. The black dashed vertical lines show $[\Delta\nu]_{0.1}$. In the left panel the dashed curves show the best-fit  $\left[C_{\ellb}(\Delta\nu)\right]_{\rm{FG}}$ corresponding to different values of $n$ indicated in the legend. The red solid curve shows the  $\left[C_{\ellb}(\Delta\nu)\right]_{\rm{FG}}$ which minimizes the residual $ [C_{\ellb}(\Delta\nu)]_{\rm{res}}$ in the range $\Delta\nu < [\Delta\nu]_{0.1}$. The dashed curves and the red solid curve in the right panel shows $[C_{\ellb}(\Delta\nu)]_{TP}$ the best-fit polynomials for $[C_{\ell}(\Delta\nu)]_T$ for the same values of $n$.}
    \label{fig:mse}
\end{figure}

The fitting procedure  is illustrated in Figure~\ref{fig:mse} where the  left panel shows the measured \clb{} for a representative $\ellb$ with $\ell = 2740$. Note that we have estimated \clb{} for the entire $\Delta \nu$ range of  $24.4\,\rm{MHz}$, which corresponds to the bandwidth of the data. The results for the entire $\Delta \nu$ range are presented in the Figure~2 of \citetalias{AE23}. We find there that various $\Delta \nu$ dependent structures  appear in \clb{} at larger $\Delta\nu$ separations $(>4\,\rm{MHz})$. These structures are possibly due to the low sampling at larger $\Delta\nu$, and also the frequency-dependence of the  PB pattern. The difficulty arises because  PF  starts to model these structures instead of  tracing  the smooth  $\Delta\nu$ dependence of the   foregrounds.  In this work, we have restricted  the $\Delta \nu$ range to $3.66 \, \rm{MHz}$, which corresponds to $150$ $\Delta \nu$  data points.  We see that  \clb{} falls  smoothly across   the $\Delta \nu$ range considered here, albeit with relatively small wriggles superimposed on this.  The right panel shows  $[C_{\ell}(\Delta\nu)]_T$ predicted for the same value of $\ell$, for which the vertical dashed line shows  $[\Delta\nu]_{0.1} = 0.56\,\rm{MHz}$ in both the panels.   Our aim is to  use the range $\Delta\nu  >  [\Delta\nu]_{0.1}$ to model  the overall smooth nature of the measured \clb{}. The left panel shows the best fit polynomials for several values of $n$. Considering $n=2$ we see that a quadratic polynomial does  not provide a very good fit. The fit improves as $n$ is increased, however the polynomial  starts to model  features from the wriggles and other rapid fluctuations if $n$ is increased beyond a certain point.  Considering $n=5,6$ and $7$, we see that these are nearly indistinguishable in the range $\Delta\nu  >  [\Delta\nu]_{0.1}$, however we see that $n=6$ provides the best fit to the measured \clb{} at $\Delta\nu  \le [\Delta\nu]_{0.1}$. The value of $\rm{MSE} $ is also minimised for $n=6$, and we have adopted this to set the order of the polynomial for this particular value of $\ellb$. We have followed the same procedure for all the $\ellb$ in our analysis, the value of $n$ was restricted to the range $0 \le n \le 10$.

PF is also expected to cause some loss of the $21$-cm signal. To quantify this loss, we consider the expected signal $[C_{\ell}(\Delta\nu)]_T$ which is shown in the right panel. Here also, we have  used   the range $\Delta\nu > [\Delta\nu]_{0.1}$ to obtain  $[C_{\ellb}(\Delta\nu)]_{TP}$ which is the  best  fit polynomial  for $[C_{\ell}(\Delta\nu)]_T$.  The best fit polynomials are shown for several values of $n$, we see that the signal loss increases as the order of the polynomial increases. Here we adopt $n=6$ which is the value used for the measured \clb{}. We see that  $[C_{\ellb}(\Delta\nu)]_{TP}$ has a peak value of  $\sim 0.04 \, \times 10^{-6} \, \rm{mK}^2 $ which is less than $10 \%$ of the peak value of $[C_{\ell}(\Delta\nu)]_T$  i.e. the signal loss is less than $10 \, \%$. In the subsequent analysis we have accounted for the signal loss using  the loss-corrected $21$-cm signal 
\begin{equation}
     \left[C_{\ellb}(\Delta\nu)\right]_{TC}= [C_{\ell}(\Delta\nu)]_{T} - [C_{\ellb}(\Delta\nu)]_{TP} \,. 
    \label{eq:TC}
\end{equation}
It may be noted  that the actual signal loss is expected to be less than the  conservative estimates used here. This is because we will generally have a combination of foregrounds and $21$-cm signal, and the foregrounds are expected to have a slower  $\Delta \nu$ variation in comparison to the $21$-cm signal. 

\subsection{Foreground modelling and removal: Gaussian Process Regression (GPR)}
\label{sec:fgmodelling_GPR}

The second approach conforms to the same idea (presented in Section~\ref{sec:fgmodelling})  that the foregrounds are expected to span the entire  $\Delta\nu$-range in contrast  to the $21$-cm signal which  largely remains localized within a small $\Delta\nu $ range. However, instead of using a polynomial (equation~\ref{eq:FGmodel}), we model the foregrounds with a Gaussian Process (GP). 

A GP is a collection of an infinite number of random variables $f(x)$ such that any finite numbers of these variables follow a joint MND. A mean $m(x)$ and a covariance function $K(x, x^{\prime})$ completely defines a GP,
\begin{equation}
    f(x) \sim \mathcal{GP} (m(x), K(x, x^{\prime})) \,.
    \label{eq:GP}
\end{equation}
In Gaussian Process Regression (GPR), the observed data $\mathbf{d}$ are assumed to be the  outcome of a GP whose mean and covariance functions are unknown. The mean and the covariance functions have forms involving hyper-parameters whose optimal values are inferred  from the data. Once the optimal values of the hyper-parameters are known, one can make predictions for unobserved data points $\mathbf{d^*}$ \citep{WR1995, RW}. The \textsc{python} library \textsc{george} \citep{george} has been used for the GPR analysis presented here.

In our case, the measured  $C_{\ellb}(\Delta \nu)$ values in the range $\Delta\nu > [\Delta\nu]_{0.1}$ (or $[\Delta\nu]_{0.4}$)  are the  observed data points $\mathbf{d}$. We model these using equation~(\ref{eq:polyfitmodel}), where $\left[C_{\ellb}(\Delta\nu)\right]_{\rm{FG}}$ and $[\rm{Noise}]$ are both assumed to be independent,  mean-zero, Gaussian random variables. We further assume the measured  $C_{\ellb}(\Delta \nu)$  to be the outcome of a GP with covariance
\begin{equation}
    \K(\dv,\dv) \equiv   K(\Delta\nu_m, \Delta\nu_n) =  k_{\rm{FG}}(\Delta\nu_m, \Delta\nu_n) + \sigma^2_N \delta_{nm}
    \label{eq:kernel_full}
\end{equation}
where $\sigma^2_N$, the noise variance, is estimated from noise-only simulations  and $ \kfg(\dv,\dv) \equiv k_{\rm{FG}}(\Delta\nu_m, \Delta\nu_n)$ is a positive semi-definite kernel which quantifies the covariance of $\left[C_{\ellb}(\Delta\nu_m)\right]_{\rm{FG}}$ and $\left[C_{\ellb}(\Delta\nu_n)\right]_{\rm{FG}}$.

The choice of the functional form of the kernel $ k_{\rm{FG}}(\Delta\nu_m, \Delta\nu_n) $ is crucial  in GPR predictions  as it encodes our assumptions (or prior knowledge) about $\left[C_{\ellb}(\Delta\nu)\right]_{\rm{FG}}$.  There are several  commonly used kernels such as squared exponential, rational quadratic  and  Mat\'ern which are stationary i.e., the kernel  depends only on  the difference $R = \mid \Delta\nu_m - \Delta\nu_n \mid$. However, in our case $\Delta \nu$ is already a frequency difference. We further find that these kernels are more sensitive to the rapid fluctuations relative to the smooth, slowly varying features in the  observed data $\mathbf{d}$. 
 
For the present analysis we have used the non-stationary polynomial kernel given by \citep{RW}
\begin{equation}
    k_{\rm{FG}}(\Delta\nu_m, \Delta\nu_n) = (\Delta\nu_m \cdot \Delta\nu_n + b )^{P}  
    \label{eq:kernel}
\end{equation}
where the constant $b$ is a  hyper-parameter and $P$, which  denotes the order of the polynomial kernel, is not a hyper-parameter and is held fixed. We have considered different values of $P$ and find  that larger values provide a better fit, however at the expense of a larger signal loss. With this in view, we have restricted ourselves to $P = 2 \, \rm{and} \, 3$  for  the entire  analysis. For a given $P$, we have used maximized the log-likelihood to estimate the optimal value of $b$ from the observed data $\mathbf{d}$.

In the final step we use GPR, with the inferred optimal hyper-parameter $b$, to make predictions for $\mathbf{d^*}$ the unobserved data points. In our case $\mathbf{d^*}$ corresponds to the foreground model predictions   $\left[C_{\ellb}(\Delta\nu)\right]_{\rm{FG}}$ for the  range  $\Delta\nu \le  [\Delta\nu]_{0.1}$ (or $[\Delta\nu]_{0.4}$). The mean and covariance of $\mathbf{d^*}$ (given the observed data $\dv$) are respectively predicted to be (see e.g. \citealt{bishop}),
\begin{align}
    \mu(\dv^* \mid \dv) &= \kfg(\dv^*,\dv) \, \K^{-1}(\dv,\dv) \, \dv  \label{eq:mean} \\ 
    \Sigma(\dv^* \mid \dv) &=  \kfg(\dv^*,\dv^*) - \kfg(\dv^*,\dv) \, \K^{-1}(\dv,\dv)  \,  \kfg(\dv,\dv^*) \,. 
    \label{eq:GPprediction}
\end{align}
Here $\mu(\dv^* \mid \dv) $ is the predicted foreground model  $\left[C_{\ellb}(\Delta\nu)\right]_{\rm{FG}}$  and  $ \Sigma(\dv^* \mid \dv)$ is the  covariance of the fitting error. 
We have added the diagonal elements of $ \Sigma(\dv^* \mid \dv)$ to the noise variance to  estimate the total error variance for $[C_{\ellb}(\Delta\nu)]_{\rm res}$. To account for  the signal loss, we have applied GPR to the expected $21$-cm signal $[C_{\ell}(\Delta\nu)]_T$ and corrected for the $21$-cm signal loss in exactly the same way  as in  Section~\ref{sec:fgmodelling}.

\subsection{Fit and residual \texorpdfstring{\clb{}}{Clb}}
\label{sec:fitandres}

We have divided the measured \clb{} into three sets, $\ell<2000$ (Set I), $2000<\ell<4000$ (Set II) and $\ell>4000$ (Set III), and  analysed them separately. For Set I, which covers  the smaller $\ell$-values,  we see that $[C_{\ell}(\Delta\nu)]_T$ de-correlates slowly with increasing $\Delta \nu$  (left panel of Figure~\ref{fig:predicted_signal}), and  $[\Delta\nu]_{0.1} \sim 1.25 \, \rm{MHz}$.  We have found that for this case the measured \clb{} contains  structures within $[\Delta\nu]_{0.1}$ which cannot be modelled  by  extrapolating the  polynomial from $\Delta\nu > [\Delta\nu]_{0.1}$. In this case, we find that it is advantageous to reduce the $\Delta\nu$ range for signal estimation and increase the $\Delta\nu$ range used for PF. Here we have used  $[\Delta\nu]_{0.4}$ for the $\ellb$ values of Set I, and $[\Delta\nu]_{0.1}$ for the $\ellb$ values in  the other two sets.  

\begin{figure}
    \centering
    \includegraphics[width=\columnwidth]{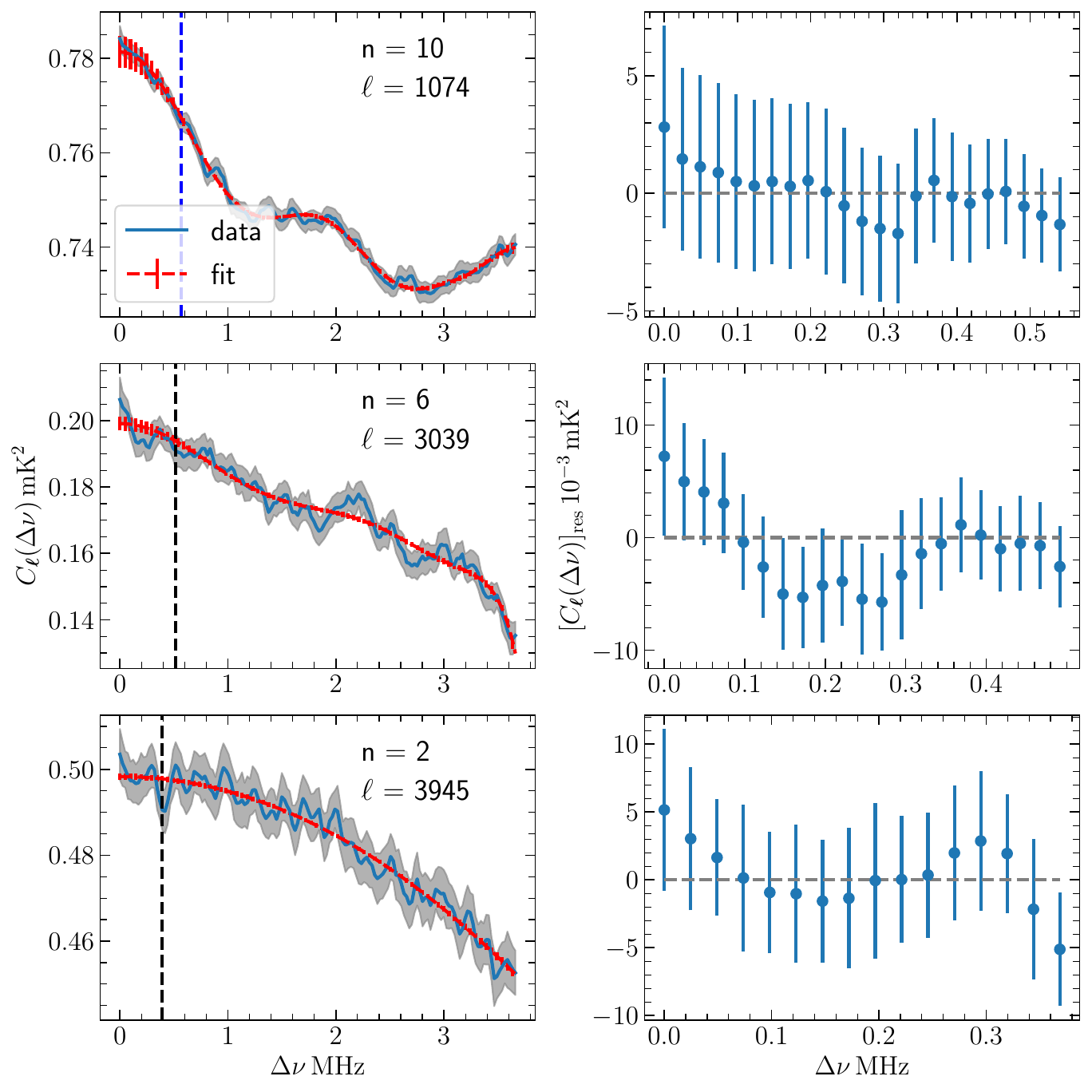}
    \caption{The blue solid lines in the left panel show the measured \clb{} with $2\sigma$ uncertainties (grey shaded region) expected from noise. The red dashed lines and the associated error bars show the best-fit polynomial foreground models and their $2\sigma$ uncertainties. The vertical dashed lines show  $[\Delta\nu]_{0.4}$ (blue) in the top panel   and $[\Delta\nu]_{0.1}$ (black) in the two bottom panels respectively. The value of $\ellb$  and $n$ (used for foreground modelling) are mentioned in the respective panels. The residual $[C_{\ellb}(\Delta \nu)]_{\rm res}$ in the range $\Delta\nu \le  [\Delta\nu]$ are shown in the right panels where  the $2\sigma$  error bars combine  the  noise and fitting  errors. }
    \label{fig:fg_fit_s1}
\end{figure}

The left panels of Figure~\ref{fig:fg_fit_s1} show the measured \clb{} (blue solid lines) along with the best-fit polynomial foreground models $\left[C_{\ellb}(\Delta\nu)\right]_{\rm{FG}}$ (red dashed lines),  considering three representative $\ellb$ values one from each set respectively.  The grey-shaded regions on the measured \clb{} show errors due to noise which were estimated using simulations as presented in \citetalias{AE23} (also in Section~\ref{sec:cylindricalps} of this paper),   whereas the red error bars on $\left[C_{\ellb}(\Delta\nu)\right]_{\rm{FG}}$ show fitting errors. All the panels here show $2\sigma$ error bars.

The right panels of Figure~\ref{fig:fg_fit_s1} show $ [C_{\ellb}(\Delta\nu)]_{\rm{res}}$, the residuals  after foreground subtraction, and the blue error bars  are the combined errors from noise and foreground modelling. 
The $\Delta \nu$ range is restricted to $\Delta \nu \le [\Delta\nu]_{0.4}$ and $[\Delta\nu]_{0.1}$  for the top panel and the two lower panels respectively.  We see that in all the panels shown here the residuals $ [C_{\ellb}(\Delta\nu)]_{\rm{res}}$ are largely consistent with zero (grey dashed lines).  While the foreground subtraction is successful at several $\ellb$ values, there  also are many $\ellb$ where the residuals are clearly not consistent with zero. In some  cases our method over-predicts   $\left[C_{\ellb}(\Delta\nu)\right]_{\rm{FG}}$ and we are left with large negative $ [C_{\ellb}(\Delta\nu)]_{\rm{res}}$, whereas we also find some other cases where  $\left[C_{\ellb}(\Delta\nu)\right]_{\rm{FG}}$  is under-predicted and we have large positive  $ [C_{\ellb}(\Delta\nu)]_{\rm{res}}$. It is necessary to identify and flag the $\ellb$ where our foreground subtraction fails and $ [C_{\ellb}(\Delta\nu)]_{\rm{res}}$ is clearly not consistent with zero.

We have assessed the effectiveness of foreground subtraction by fitting  the residual $ [C_{\ellb}(\Delta\nu)]_{\rm{res}}$ using our model prediction for the expected signal $ \left[C_{\ell}(\Delta\nu)\right]_T$ (equation~\ref{eq:cl_Pk_sinc})  with ${\bf A} \equiv \left[\Omega_{\HI} b_{\HI}\right]^{2}$ as a free parameter. The quantity ${\bf A}$ effectively quantifies the amplitude of  $ [C_{\ellb}(\Delta\nu)]_{\rm{res}}$, with ${\bf dA}$ denoting the predicted uncertainties. We have used the criteria $\mid {\bf A} \mid/{\bf dA} > 2$ to identify the $\ellb$ values where foreground subtraction fails,  which are then flagged. We finally have $8, 33 \, \rm{and} \,20$ unflagged $\ellb$ values for Sets I, II and III, respectively. Figures~\ref{fig:fg_fit_s1a} to \ref{fig:model_s3} of Appendix~\ref{sec:sec2n3} show the measured \clb{}, the corresponding foreground model  $\left[C_{\ellb}(\Delta\nu)\right]_{\rm{FG}}$ and the  residual  $[C_{\ellb}(\Delta\nu)]_{\rm{res}}$ for all the $\ellb$ values which were accepted for the subsequent analysis. A visual inspection shows  these  $[C_{\ellb}(\Delta\nu)]_{\rm{res}}$ values to be largely within the $2 \sigma$ error bars. We have  also considered   the flagging criteria ($\mid {\bf A} \mid/{\bf dA} >1$)  and  ($\mid {\bf A}\mid / {\bf dA} >3$ and $5$ ) to investigate what happens if the cut is tightened or relaxed. We find that a few more $\ellb$ values are flagged if the cut is tightened, whereas we pick up some $\ellb$ values having residual foregrounds if the cut is relaxed. Either way, the final conclusions are not much changed, and we have used $\mid {\bf A} \mid/{\bf dA} > 2$ which provides the best results.

\begin{figure}
    \centering
    \includegraphics[width=\columnwidth]{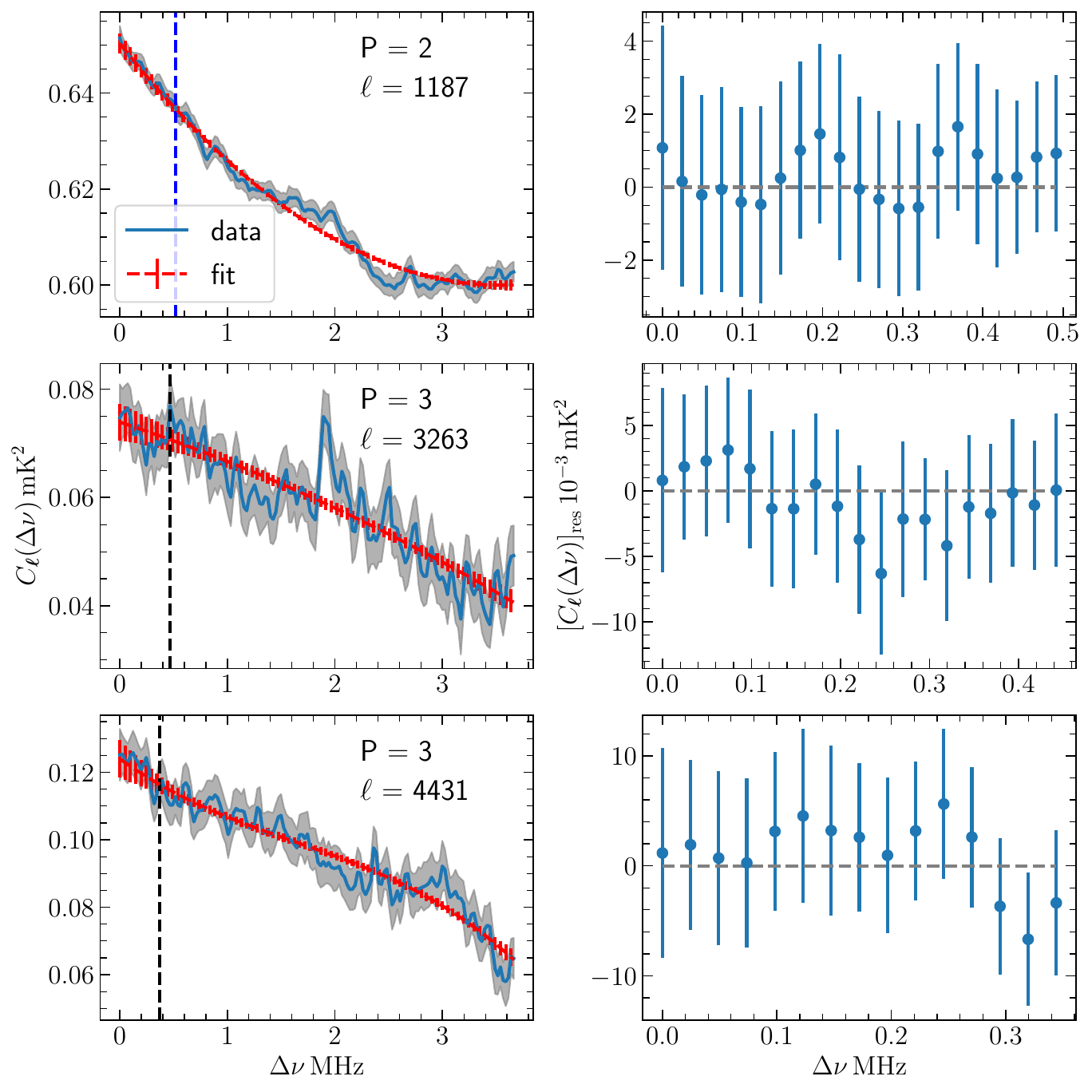}
    \caption{Same as Figure~\ref{fig:fg_fit_s1} but for GPR. The quoted values of $P$ denote the orders of the polynomial covariance function. }
    \label{fig:fg_fit_s1_GPR}
\end{figure}

Figure~\ref{fig:fg_fit_s1_GPR} is the same as  Figure~\ref{fig:fg_fit_s1}, except that we have used GPR instead of PF for foreground modelling. The flagging here has also been carried out in the same way as for  PF. For GPR, visual inspection reveals a further $4$ $\ellb$ grid points in Set III where $[C_{\ellb}(\Delta\nu)]_{\rm{res}}$  has strong  oscillatory features which can be attributed to residual point source foregrounds. In addition to the flagging criteria described earlier, we have also flagged these $4$ visually identified $\ellb$ grids. We finally get $2, 21$ and $13$ unflagged $\ellb$ values in GPR analysis for Sets I, II and III, respectively. Figures~\ref{fig:fg_fit_GPR} and \ref{fig:model_GPR} of Appendix~\ref{sec:sec2n3} show the measured \clb{}, the corresponding foreground model  $\left[C_{\ellb}(\Delta\nu)\right]_{\rm{FG}}$ and the  residual  $[C_{\ellb}(\Delta\nu)]_{\rm{res}}$ for all the $\ellb$ values which were accepted for the subsequent analysis. We visually assess that these $[C_{\ellb}(\Delta\nu)]_{\rm{res}}$ values are largely within the $2 \sigma$ error bars.

We note a few features common to both PF and GPR analysis presented here. Considering the foreground fits, both the methods show quite reasonable fits to the measured \clb{}, and the residuals $ [C_{\ellb}(\Delta\nu)]_{\rm{res}}$ are largely within $2\sigma$ error bounds. Strictly speaking, we could interpret the residuals as being consistent with the noise predictions if  the $ [C_{\ellb}(\Delta\nu)]_{\rm{res}}$ values were  randomly distributed around zero. However, we find that the residuals are not random but exhibit  correlations which extend over several adjacent $\Delta \nu$ values. Here we note that this can arise from  uncertainties  in  foreground modelling  which  are, in general, predicted to be correlated (e.g. equation~\ref{eq:GPprediction}).  Low-level residual foregrounds can also lead to such correlations. In order to keep the treatment simple, for the present analysis we have ignored these correlations and we have treated the errors at each $ [C_{\ellb}(\Delta\nu)]_{\rm{res}}$  as being independent.

\section{The Cylindrical PS}
\label{sec:cylindricalps}

The cylindrical PS $P(k_{\perp}, k_{\parallel})$  of the $21$-cm signal is the Fourier transform of the MAPS \cl{} along $\Delta\nu$ \citep{KD07}
\begin{equation}
    P(k_{\perp}, k_{\parallel})= r^2\,r^{\prime} \int_{-\infty}^{\infty}  d (\Delta \nu) \, \mathrm{e}^{-i  k_{\parallel} r^{\prime} \Delta  \nu}\, C_{\ell}(\Delta \nu) \,.
\label{eq:cl_Pk}
\end{equation}
Following \citetalias{AE23} we model the measured $C_{\ell}(\Delta \nu)$ as
\begin{equation}
    C_{\ell}(\Delta\nu_n)=  \sum_{m} \textbf{A}_{nm} P(k_{\perp}, k_{\parallel m})
   + [\textrm{Noise}]_{n}
\label{eq:CL_data}
\end{equation}
where $\textbf{A}$ contains the Fourier transform coefficients and $[\textrm{Noise}]_{n}$ is the noise in each estimated $C_{\ell}(\Delta\nu_n)$. 
The maximum likelihood estimate of $P(k_{\perp}, k_{\parallel m})$ is given by, 
\begin{equation}
    P(k_{\perp}, k_{\parallel m})   =  \sum_n \left[ \left(\textbf{A} ^{\dagger} \textbf{N}^{-1} \textbf{A}\right)^{-1} \textbf{A}^{\dagger} \textbf{N}^{-1} \right]_{mn} \, C_{\ell}(\Delta\nu_n)
    \label{eq:MLE_PS}
\end{equation}
where $\textbf{N}$ is the noise covariance matrix. 

The noise covariance estimate is detailed in \citetalias{AE23} which we briefly restate here. We have simulated multiple ($50$) realizations of visibility data having zero mean and standard deviation  $ \sigma_{N}=0.43 \, \rm{Jy}$, which is the visibility r.m.s present in the actual data. We identically analyse the simulated data using the TGE (equation~\ref{eq:crossmaps}) to estimate $\textbf{N}$. However, the true noise level in the data is $\sim 4.77$ times larger than what we obtain from the system `noise-only' simulations (\citetalias{AE23}), and the noise levels are scaled up with this factor for the present work.

We have binned the foreground subtracted  residual  $[C_{\ellb}(\Delta\nu)]_{\rm{res}}$ at different grid points $\ellb_g$ into  equally separated annular bins in the $uv$-plane. We have calculated the bin-averaged values  $C_{\ell_a}(\Delta\nu_n) = \frac{\sum_g w_g [C_{\ellb}(\Delta\nu)]_{\rm{res}}}{\sum_g w_g}$ at the bin-averaged multipoles $\ell_a = \frac{\sum_g w_g \ell_g}{\sum_g w_g}$, where the weight $w_g$ of the grid points are taken to be unity implying that all grids are equally weighted. The $\ell_a$  values here span the range $1000 \lesssim \ell_a \lesssim 5800$ with $5$ $\ell$ bins, this  corresponds to the  $k_{\perp}$ range  $0.18 \lesssim k_{\perp} \lesssim 1.0\,\rm {Mpc}^{-1}$ for  the estimated $P(k_{\perp}, k_{\parallel})$. The $\Delta \nu$ extent available for $21$-cm signal estimation is different  for each  $\ellb$. However, for estimating the cylindrical PS we have used a fixed range of  $0.488 \, \rm{MHz}$ which corresponds to $N_E = 20$ frequency separations $\Delta \nu$ for all the bins. As a consequence, the estimated $P(k_{\perp}, k_{\parallel})$ all span the same  $k_{\parallel}$  range of $0\le k_{\parallel} \le 13.1 \, \rm{Mpc}^{-1}$ with the resolution $\Delta k_{\parallel} = 0.69 \, \rm{Mpc}^{-1}$.  The actual available $\Delta \nu$ range is larger than  $0.488 \, \rm{MHz}$ for some of the small $\ellb$ for which this assumption leads to some additional $21$-cm signal loss. The actual available range is somewhat smaller than $0.488 \, \rm{MHz}$ for the large $\ellb$,  in which case we end up introducing excess noise into the analysis. Note that the exact available $\Delta \nu$ ranges are shown for all the $\ellb$ in  Appendix~\ref{sec:sec2n3}. We further note that the cylindrical PS has not been used for any of the final quantitative outcomes quoted in this paper. Here the cylindrical PS has primarily been used for a visual representation of the residual data and for exploring the noise statistics. It may also be noted that we have not corrected $P(k_{\perp}, k_{\parallel})$ for the signal loss associated with the foreground removal.  We do not expect these simplifying assumptions to have a severe impact for the two issues under consideration in this section. 

\begin{figure*}
    \centering
    \includegraphics[width=\textwidth]{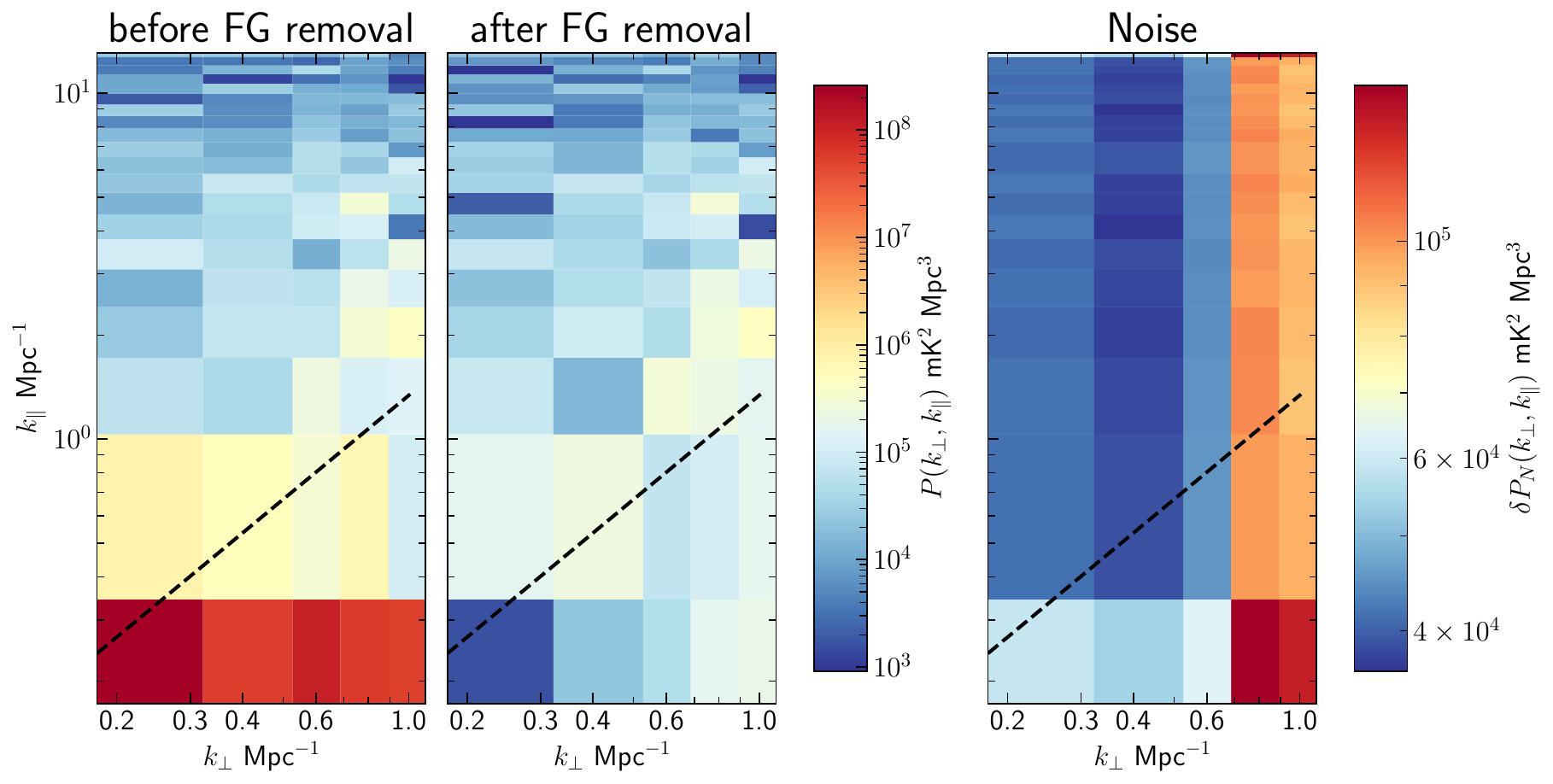}
    \caption{ The first two panels show the cylindrical PS $\mid P(k_{\perp}, k_{\parallel}) \mid$ before and after foreground removal (using PF). The third panel shows the $1\sigma$ statistical fluctuation  $\delta P_N(k_{\perp},k_{\parallel})$ in the estimated $P(k_{\perp}, k_{\parallel})$. The black dashed line shows the predicted boundary of the foreground wedge $[k_{\parallel}]_{H}$.  }
    \label{fig:cylps}
\end{figure*}

Figure~\ref{fig:cylps} shows $\mid P(k_{\perp},k_{\parallel}) \mid$ before and after foreground removal using PF in the left and the middle panels respectively. The cylindrical PS are visually very similar for GPR analysis, and  we do not explicitly show them here. The rightmost panel shows  $\delta P_N(k_{\perp},k_{\parallel})$ the expected r.m.s. statistical fluctuation due to noise. We note that $\delta P_N(k_{\perp},k_{\parallel})$  is obtained by applying the maximum likelihood estimator (MLE; equation~\ref{eq:MLE_PS}) on the \cl{} obtained from   multiple realizations of the noise only simulations mentioned earlier. The black dashed lines show the predicted location of the foreground wedge boundary $  [k_{\parallel}]_{H} = (r/r^{\prime}\nu_{c}) k_{\perp} $.   The left panel  shows  that  before  foreground subtraction $\mid P(k_{\perp}, k_{\parallel}) \mid$ has a dynamic range of $\sim10^6$ starting from $\sim10^8  \, \rm{mK}^2\,\rm{Mpc}^3$ at $k_\parallel = 0$ to $\sim10^2  \, \rm{mK}^2\,\rm{Mpc}^3$ at the higher $k_\parallel$. The most affected LoS mode is $k_{\parallel} = 0$ which corresponds to the DC of the signal. Note that we have shifted the $k_{\parallel} = 0$ point slightly to place  it in a log scale. We notice that the major contribution of the power is within $[k_{\parallel}]_{H}$. A significant foreground leakage is also noticeable up to $k_{\parallel} \sim 1 \,\rm{Mpc}^{-1}$. The $\mid P(k_{\perp},k_{\parallel}) \mid$ values beyond $k_{\parallel} \sim 1 \,\rm{Mpc}^{-1}$  appear to be  comparatively foreground-free. 

Considering the middle panel we see that the $\mid P(k_{\perp},k_{\parallel}) \mid$ values are found to lie within  $10^2 - 10^6  \, \rm{mK}^2\,\rm{Mpc}^3$. The $\mid P(k_{\perp},k_{\parallel}) \mid$ values at  $k_{\parallel} < 1 \,\rm{Mpc}^{-1}$ have an amplitude $\sim10^5  \, \rm{mK}^2\,\rm{Mpc}^3$. We see that the smooth foreground components which appear at low $k_{\parallel}$ have been  successfully subtracted out  from the data. In the last two $k_{\perp}$ bins, we see some residual foregrounds at large $k_{\parallel} \, (\geq 2\,\rm{Mpc}^{-1})$. In these larger $k_{\perp}$ modes, the noise is found to be high (right panel), and we could not separate a  rapidly varying foreground component (or possibly unknown systematics) from the noisy data.

After foreground subtraction, we can use the entire $(k_{\perp}, k_{\parallel})$ space for constraining the $21$-cm PS. However, it is necessary to ensure that the measured $P(k_{\perp}, k_{\parallel})$ values are either strictly positive, or be of either signs with the negative values being consistent with the expected noise level. Due to the non-uniform baseline sampling in the $uv$-plane and also in $\Delta \nu$, we do not expect the  noise  level to be uniform across the $(k_{\perp}, k_{\parallel})$ plane. We account for this by considering the quantity   $X$  which is defined as \citep{Pal20}
\begin{equation}
    X=\frac{P(k_{\perp},k_{\parallel})}{\delta P_{N}(k_{\perp},\,k_{\parallel})}.
    \label{eq:xstat}
\end{equation}
The distribution of $X$ is expected to be symmetric with mean $\mu = 0$ and standard deviation $\sigma_{Est}=1$  if the estimated $P(k_{\perp},k_{\parallel})$ are completely due to uncorrelated Gaussian random noise in the measured visibilities. Figure 5 of \citetalias{AE23} shows the histogram of $X$ prior to foreground subtraction  considering only the   $(k_{\perp},k_{\parallel})$ modes within the `$21$-cm window' (TW). This was found to be largely symmetric around $\mu = 0.61$  but with $\sigma_{Est}=4.77$. The small, positive mean $(\mu >0)$ was interpreted as indicating the presence of some low-level foreground leakage into the TW, whereas the relatively large $\sigma_{Est}$  indicates that the actual $\delta P_N(k_{\perp},k_{\parallel})$ are larger than those expected from system noise alone. As noted in \citetalias{AE23},  possible sources for this excess noise include artefacts due to imperfect calibration, low-level residual RFIs, inaccuracy in point source removal,  and various other factors which are not well known at present \citep{mertens20, Gan2022}. As mentioned earlier, we have scaled up the predictions from the system noise-only simulations by the factor $\sigma_{Est}=4.77$ to account for this excess noise. It is important to note that this scaling factor plays a crucial role in interpreting the statistical significance of the results presented here and also in  \citetalias{AE23} as most of the PS measurements would be well above (or below) the expected noise level if this scaling were not applied.

\begin{figure*}
    \centering
    \includegraphics[width=\textwidth]{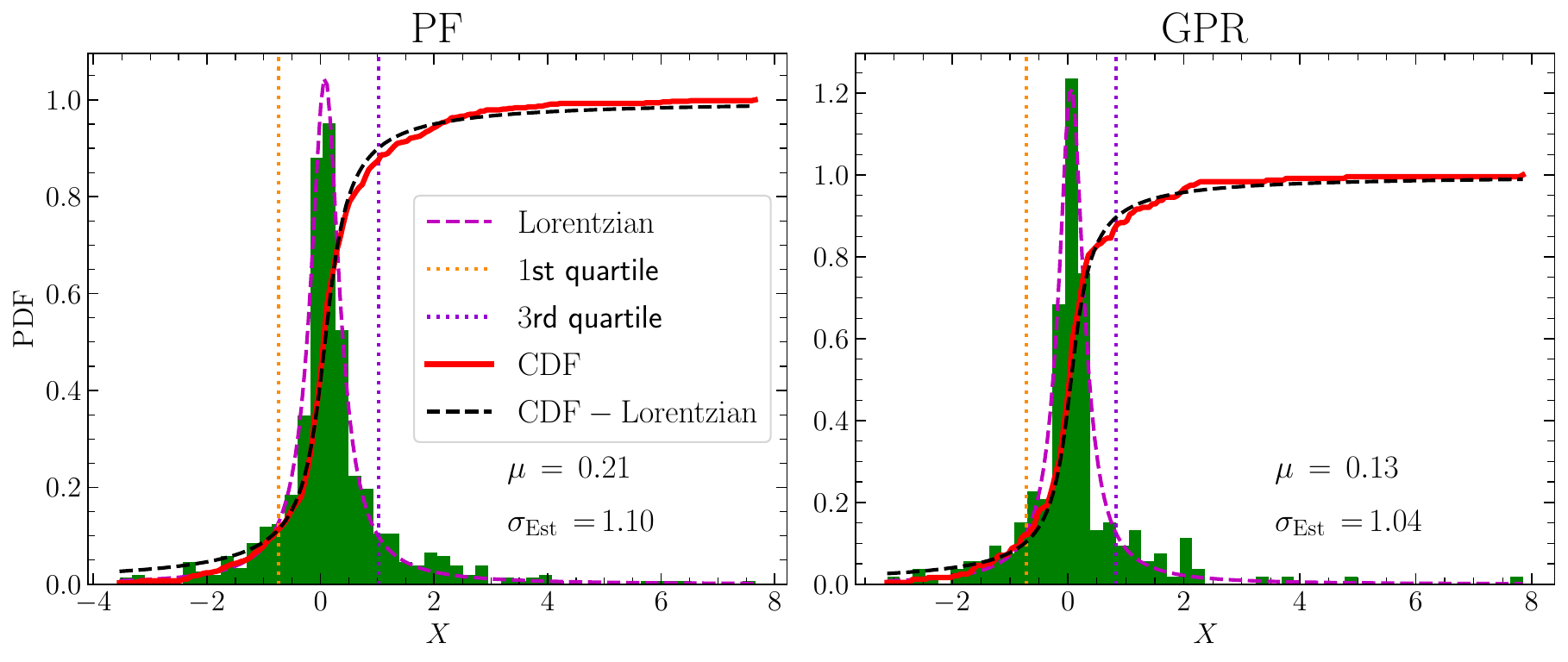}
    \caption{This shows the probability density function (PDF; green vertical bars ) and the   cumulative distribution function (CDF; red solid line) for  $X = \frac{P(k_{\perp}, \,k_{\parallel})} {\delta P_{N}(k_{\perp},\,k_{\parallel})}$. Lorentzian fits of the PDF and the CDF are shown by the magenta and black dashed lines respectively. The orange and violet vertical lines show the first and the third quartile of the best-fit Lorentzian distribution. The mean $(\mu)$ and the standard deviation  $(\sigma)$ of $X$ are annotated.}
    \label{fig:noisestat}
\end{figure*}

Figure~\ref{fig:noisestat} of this paper  shows the histogram of $X$ after foreground subtraction using  PF (left) and GPR (right) considering all the available $(k_{\perp},k_{\parallel})$ modes.  Note that in addition to the scaled system noise, ${\delta P_{N}(k_{\perp},\,k_{\parallel})}$ now also has a contribution from the uncertainties in foreground modelling.  We find $\mu=0.21$ and $\sigma_{Est}=1.1$ for  PF  and $\mu = 0.13$ and $\sigma_{Est} = 1.04$ for  GPR, whereas ideally we expect $\mu=0$ if the foregrounds have been perfectly subtracted and $\sigma_{Est}=1$ if our predictions for the expected statistical fluctuations are correct. The small, positive value of $\mu$ indicates that some low-level foregrounds possibly still remain in the residuals.  However, note  that in both cases the value of $\mu$ is smaller compared to that before foreground subtraction. This indicates an overall reduction in the level of foreground contamination for $21$-cm signal estimation as compared to \citetalias{AE23}. The fact that we obtain  $\sigma_{Est} \approx 1$ for both PF and GPR roughly validates our error predictions including our treatment of the uncertainties due to foreground modelling.  

For both PF and GPR, we see that the probability density function (PDF) of $X$ is largely  symmetric around  a small positive mean value.  We find that the  bulk $(\sim90-95\%)$ of  the  $X$ values  lie in the central  region $ \mid X \mid \le 2$. We find a few samples $(\sim1\%)$ at larger values of $X$ (e.g. $X>4$), which are trace amounts of unsubtracted residual foregrounds and can be considered outliers of the distribution. The fact that we do not see any significant outliers indicates that foregrounds have largely been subtracted from the data. Similar to \citetalias{AE23}, we find that  PDFs  are not well described by a Gaussian, whereas a Lorentzian distribution (also known as Cauchy distribution) represents the statistics better. The magenta dashed lines in Figure~\ref{fig:noisestat} show the best-fit Lorentzian PDFs 
\begin{equation}
    \rho( x ) = \frac{1}{\mathrm{\pi} \gamma} \left[ \frac{\gamma^2}{(x-x_0)^2 + x_0^2} \right]
    \label{eq:Lorentzianpdf}
\end{equation}
with $x_0$ and $\gamma$  respectively denoting the peak location  and the spread (half-width at half-maximum) of the distribution.  We find $x_0 = 0.08$ and $\gamma = 0.30$ for PF, and $x_0 = 0.06$ and $\gamma = 0.26$ for GPR. Figure~\ref{fig:noisestat} also shows the cumulative distribution function (CDF) of $X$, along with the CDF predicted by the best-fit Lorentzian distribution. We see that for both the foreground subtraction methods, the CDF of $X$ and the corresponding best-fit Lorentzian CDF are in close agreement.

The quantity of our interest is the statistical uncertainties in the estimated $P(k_{\perp},k_{\parallel})$. We have found that $X$ is well represented by a Lorentzian distribution, but it is not guaranteed that the statistical fluctuations  in the estimated $P(k_{\perp},k_{\parallel})$  will also be described by the same. \cite{Wilensky23} recently found that this  is  likely  to follow a Laplacian distribution, and converge to a Gaussian in most situations. We also note that the standard deviation is ill-defined for the  Lorentzian distribution in our analysis. Considering this uncertainty, we have followed the standard practice of quoting the $2\sigma$ error bars throughout the analysis. However,  it is questionable if these error bars actually represent the  $95\%$ confidence intervals. 

Following \citetalias{AE23}, for the subsequent analysis we have further scaled the error  estimates with  $\sigma_{Est}$. Note that this additional scaling  leads to a very small increase in the error bars as $\sigma_{Est} \approx 1$ for both PF and GPR.

\section{The Spherical PS}
\label{sec:BMLE}
We can think of the  foreground subtracted  residual   $[C_{\ellb}(\Delta\nu)]_{\rm{res}}$ in terms of two components 
\begin{equation}
    \left[C_{\ellb}(\Delta\nu)\right]_{\rm{res}} = \left[C_{\ell}(\Delta\nu)\right]_{T} + \left[C_{\ellb}(\Delta\nu)\right]_{R}\, .
    \label{eq:Cmodel}
\end{equation}
Here $[C_{\ell}(\Delta\nu)]_{T}$ is the  spatially isotropic $21$-cm signal and $[C_{\ellb}(\Delta\nu)]_{R}$ refers to the anisotropic components of $C_{\ellb}(\Delta \nu)$ which is ideally expected to be  within  the noise level.  Un-subtracted foregrounds, if present, will also contribute to $[C_{\ellb}(\Delta\nu)]_{R}$. 

Here  the  $21$-cm signal is modelled as 
 \begin{equation}
     \left[C_{\ell_a}(\Delta\nu_n)\right]_{T} =   \sum_{i} B_i(a , n) \left[ P(k_i) \right]_{T} \label{eq:CTPK}
 \end{equation}
where $[P(k_i)]_T$ refers to the value of the spherical $21$-cm PS in the $i$-th bin and  $B_i(a , n) = \sum_{m} A_{nm}$ is  summed   over the  $(k_{\perp a},k_{\parallel m})$ modes  within  this bin. We have applied a maximum likelihood estimator (MLE; \citetalias{AE23} ) to directly determine  PS $P(k)$ from the measured $[C_{\ellb}(\Delta\nu)]_{\rm{res}}$.

\begin{figure}
    \centering
    \includegraphics[width=\columnwidth]{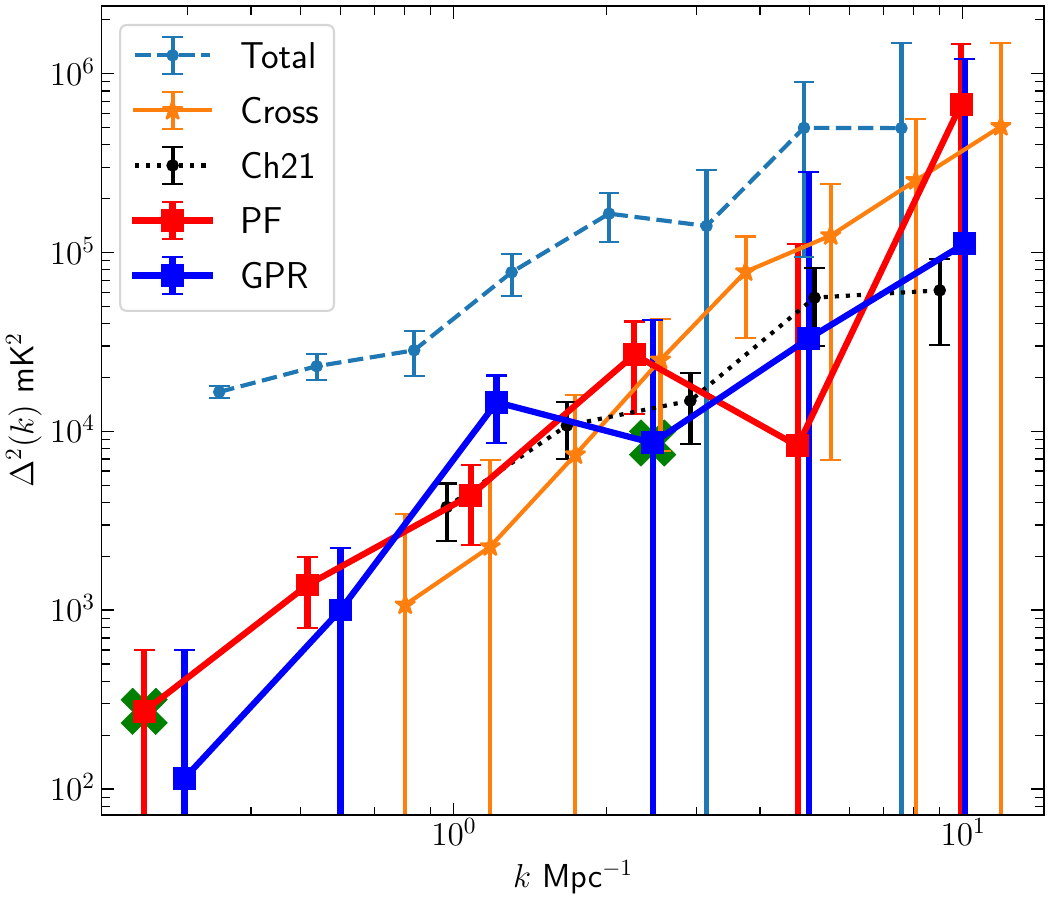}
    \caption{ The mean squared brightness temperature fluctuations $\Delta^2(k) $ and the associated $2 \sigma$ errors are shown. The red and blue squares show the values obtained in this work using PF and GPR, respectively. The green crosses $(\textbf{X})$ indicate negative values. The  light-blue (dashed), orange (solid) and  black (dotted) curves present the results of \citetalias{P22}, \citetalias{AE23}  and \citetalias{Ch21}  respectively. } 
    \label{fig:pssph}
\end{figure}

\begin{table}
    \centering
    \caption{The mean squared brightness temperature fluctuations $\Delta^2(k)$, their errors $\sigma$, signal-to-noise ratio (SNR), $2\,\sigma$ upper limits   $\Delta_{\rm UL}^{2}(k) = \Delta^{2}(k)+2 \, \sigma$ and $[\Omega_{\HI}b_{\HI}]_{\rm UL}$ values are tabulated for PF.}
        \begin{tabular}{cccccc}
        \hline
        \hline
        $k$ & $\Delta^2(k)$ & $1\sigma$ & SNR & $\Delta_{\rm UL}^{2}(k)$ & $[\Omega_{\HI}b_{\HI}]_{\rm UL}$ \\
        $\rm{Mpc}^{-1}$ & $\rm{mK}^2$ & $\rm{mK}^2$ & & $\rm{mK}^2$  & \\
        \hline
        $0.247$ &  $-(16.50)^2$ & $(12.78)^2$ & $-1.67$ & $(18.07)^2$ & $0.036$ \\
        $0.517$ &  $(37.20)^2$ & $(17.21)^2$ & $4.67$ & $(44.46)^2$ & $0.064$ \\
        $1.082$ &  $(66.35)^2$ & $(32.26)^2$ & $4.23$ & $(80.52)^2$ & $0.089$ \\
        $2.266$ &  $(163.84)^2$ & $(84.59)^2$ & $3.75$ & $(202.86)^2$ & $0.180$ \\
        $4.744$ &  $(91.06)^2$ & $(226.59)^2$ & $0.16$ & $(333.14)^2$ & $0.245$ \\
        $9.931$ &  $(820.54)^2$ & $(626.28)^2$ & $1.72$ & $(1207.37)^2$ & $0.759$ \\
        \hline
    \end{tabular}
    \label{tab:ul_MLE}
\end{table}

\begin{table}
    \centering
    \caption{Same as Table~\ref{tab:ul_GPR} but for GPR.}
        \begin{tabular}{cccccc}
        \hline
        \hline
        $k$ & $\Delta^2(k)$ & $1\sigma$ & SNR & $\Delta_{\rm UL}^{2}(k)$ & $[\Omega_{\HI}b_{\HI}]_{\rm UL}$ \\
        $\rm{Mpc}^{-1}$ & $\rm{mK}^2$ & $\rm{mK}^2$ & & $\rm{mK}^2$  & \\
        \hline
        $0.296$ &  $(10.72)^2$ & $(15.61)^2$ & $0.47$ & $(24.54)^2$ & $0.045$ \\
        $0.600$ &  $(31.68)^2$ & $(24.71)^2$ & $1.64$ & $(47.17)^2$ & $0.064$ \\
        $1.215$ &  $(120.66)^2$ & $(54.65)^2$ & $4.87$ & $(143.29)^2$ & $0.153$ \\
        $2.461$ &  $-(92.83)^2$ & $(128.75)^2$ & $-0.52$ & $(182.08)^2$ & $0.158$ \\
        $4.985$ &  $(181.84)^2$ & $(353.91)^2$ & $0.26$ & $(532.51)^2$ & $0.387$ \\
        $10.098$ &  $(334.80)^2$ & $(741.99)^2$ & $0.20$ & $(1101.45)^2$ & $0.681$ \\
        \hline
    \end{tabular}
    \label{tab:ul_GPR}
\end{table}

We have used all the available $[C_{\ellb}(\Delta\nu)]_{\rm{res}}$ values to obtain the the model parameters $P(k)$ at $6$ spherical $k$ bins spanning the range  $0.247 < k < 9.931 \, \rm{Mpc}^{-1}$. We find that the goodness-of-fit parameter (reduced-$\chi^2$) has a value of $1.14$ with $629$ degrees of freedom. Although this is a slightly poor fit for $[C_{\ellb}(\Delta\nu)]_{\rm{res}}$, we have used the  best fit  $P(k)$ values to calculate the mean squared brightness temperature $\Delta^2(k)\equiv {k^{3}}P(k)/{2\mathrm{\pi}^{2}}$. The red and blue curves in Figure~\ref{fig:pssph} show   $\mid \Delta^2(k) \mid$ along with the corresponding $2\sigma$ error bars for PF and GPR respectively. Note that we have marked the negative $\Delta^2(k)$ values with a (green) cross $(\textbf{X})$. Considering  PF, we see that the $\Delta^2(k)$ values (red squares) are positive in all the $k$ bins except the first $k$ bin $(k = 0.247 \, \rm{Mpc}^{-1})$. For GPR,  $\Delta^2(k)$  is found to be negative in the fourth bin $(k = 2.461 \, \rm{Mpc}^{-1})$.  While we expect  $\Delta^2(k)$   for the $21$-cm signal to  be positive, the  values inferred from  observations can be negative due to statistical fluctuations. We see that the negative $\Delta^2(k)$ values  are  all within $0\pm 2\sigma$, and we interpret  these measurements as being consistent with zero with the negative values  arising from statistical fluctuations. The  values of $\Delta^2(k)$, $\sigma$  and the SNR ($=\Delta^2(k)/\sigma$) at different $k$-bins  are presented  in Tables~\ref{tab:ul_MLE} and ~\ref{tab:ul_GPR} for PF and GPR respectively. Note that the correction due to signal loss, which we will discuss shortly, is considered for all the values quoted here. The  $2\sigma$  upper limits $\Delta_{\rm UL}^{2}(k) = \Delta^{2}(k) + 2\sigma$ obtained from the two methods are also presented in their corresponding tables. Note that for the negative $\Delta^2(k)$ values,  we have conservatively taken $\Delta^2(k) = 0$ and quoted the $2 \sigma$ values as the upper limits. The tightest constraint is found to be  $\Delta_{\rm UL}^2(k)\leq (18.07)^2 \, \rm{mK}^2$ at $k = 0.247 \, \rm{Mpc}^{-1}$ from PF.

\begin{table}
\centering
\caption{The upper limits from  $21$-cm IM experiments using this uGMRT Band 3 data. The \omb{}$_{\rm UL}$ values quoted inside the parentheses (...) are obtained when a single, $k$-independent, value of \omb{} is directly constrained from \cl{} (Section~\ref{sec:omb}).} 
    \begin{tabular}{cccccc}
        \hline
        \hline
        Works & $z$ & k & $[\Delta^2(k)]_{\rm UL}$ &
        \omb{}$_{\rm UL}$  \\
        & & $\rm{Mpc}^{-1}$ & $\rm{mK}^2$& \\
        \hline
        \multirow{4}{*}{\citetalias{Ch21}} & 1.96 & 0.99 & $(58.57)^2$ & $0.09$\\
         & 2.19 & 0.97 & $(61.49)^2$ & $0.11$\\
         & 2.62 & 0.95 & $(60.89)^2$ & $0.12$\\
         & 3.58 & 0.99 & $(105.85)^2$ &$0.24$\\
        \hline
        \citetalias{P22} & 2.28 & 0.35 & $(133.97)^{2}$ & $0.23$\\
        \citetalias{AE23} & 2.28 & 0.80 & $(58.67)^{2}$  & $0.072 \, (0.061)$ \\
        \hline
        \rm{Present work} &  &  &  & \\        
        \rm{PF} & \multirow{2}{*}{$2.28$} & $0.25$ & $(18.07)^2$ & $0.036 \, (\mathbf{0.022})$\\
        \rm{GPR} &  & $0.30$ & $(24.54)^2$ & $0.045 \, (\mathbf{0.031})$\\
        \hline
    \end{tabular}
    \label{tab:results_IM}
\end{table}

The results obtained from  previous works with the same observational data are also shown in Figure~\ref{fig:pssph}. \citetalias{P22} (blue-dashed) used the Total TGE whereas \citetalias{AE23} (solid orange) has used the Cross TGE which was also adopted for the present  work. We also present the results from \citetalias{Ch21} (black dotted line)  who have estimated the $21$-cm PS in delay space using an $8$ MHz bandwidth data taken from the same observations at $\nu_c = 445 \,\rm{MHz} \, (z = 2.19)$. All of these works have used foreground avoidance and the PS estimates are restricted to $k> 0.35, 0.80$ and $1 \, \rm{Mpc}^{-1}$ respectively due to  the presence of the foreground wedge.  

In the present analysis we are able to foray deeper into the  small $k$ range extending up to $k = 0.247 \, \rm{Mpc}^{-1}$. While our results are consistent with the earlier estimates of \citetalias{AE23} and \citetalias{Ch21} at larger $k$ where there is an overlap, we obtain a tighter upper limit of $\Delta_{\rm UL}^2(k)\leq (18.07)^2 \, \rm{mK}^2$ at the smallest $k$ bin of $k = 0.247 \, \rm{Mpc}^{-1}$. Note that a comparison of all the upper limits obtained from the same observation is presented in Table~\ref{tab:results_IM}.  

\begin{figure}
    \centering
    \includegraphics[width=\columnwidth]{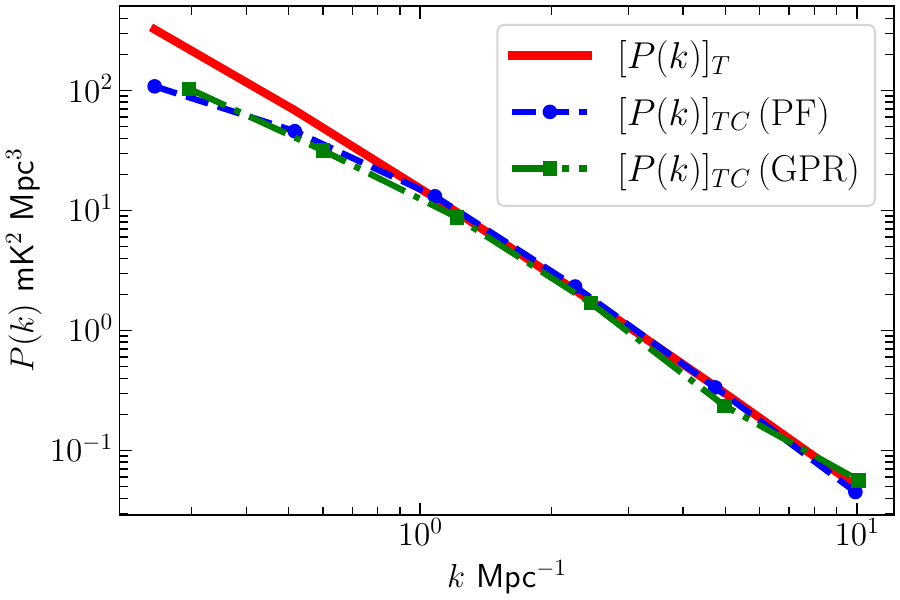}
    \caption{Correction in the spherical PS $P(k)$  to take into account the signal loss due to the foreground removal algorithms. The red line shows the  predicted $21$-cm PS, whereas the blue and green lines  show the loss-corrected $21$-cm PS from PF and GPR, respectively.} 
    \label{fig:correction}
\end{figure}

The results presented above have all been corrected for possible $21$-cm signal loss due to foreground subtraction. To estimate the correction factor we have calculated the spherically binned  PS using both $[C_{\ell}(\Delta\nu)]_T$ (equation~\ref{eq:cl_Pk_sinc}) and $[C_{\ell}(\Delta\nu)]_{TC}$ (equation~\ref{eq:TC}). The analysis has been done identical to the analysis of the measured \clb{}. Figure~\ref{fig:correction} shows $[P(k)]_T$ and $[P(k)]_{TC}$ estimated from $[C_{\ell}(\Delta\nu)]_T$ and $[C_{\ell}(\Delta\nu)]_{TC}$ respectively. We have multiplied the estimated  $\Delta_{\rm UL}^2(k) $ at each $k$ bin with the corresponding correction factor $[P(k)]_T/[P(k)]_{TC}$  to account for the signal loss. We see that the correction factor is very close to unity for all the $k$ bins expect for the smallest two bins. Considering PF (GPR), the correction factor has values  $\sim3.0$ ($\sim2.2$)  and $\sim1.5$ ($\sim 1.5$) for the first two $k$ bins respectively. The rather large signal loss in the lowest $k$-bin can be attributed to the fact that we have used $[\Delta \nu]_{0.4}$ for Set I which has the smallest $k_{\perp}$ modes.

The upper limits on $\Delta^2(k)$ allow us to constrain the cosmological \HI{} abundance parameter $[\Omega_{\HI}b_{\HI}]$. The assumption that the \HI{} distribution traces the underlying matter distribution allows us to express  $P_{T}(\mathbf{k})$ in terms of  $P^s_m(\mathbf{k})$  the matter PS in redshift space through the relation (equation 17 of \citetalias{AE23})
\begin{equation}
    P_{T}(\mathbf{k}) = \left[\Omega_{\HI} b_{\HI}\right]^{2} \bar{T}^{2} P^s_{m}(\mathbf{k})
    \label{eq:pT}
\end{equation}
with the mean brightness temperature $\bar{T}$  given in equation~(\ref{eq:tbar}). Using the estimated $\Delta_{\rm UL}^{2}(k)$ values we place $2 \sigma$ upper limits   $[\Omega_{\HI}b_{\HI}]_{\rm UL}$ which are also presented in Tables~\ref{tab:ul_MLE} and ~\ref{tab:ul_GPR}. The best constraint is obtained using PF where we find $[\Omega_{\HI}b_{\HI}]_{\rm UL} \le  0.036$ from the first bin $k = 0.247 \, \textrm{Mpc}^{-1}$.

\section{Constraining \texorpdfstring{\omb{}}{Omb}}
\label{sec:omb}

Here we use the foreground subtracted residuals  $[C_{\ellb}(\Delta\nu)]_{\rm{res}}$ to directly measure $[\Omega_{\HI}b_{\HI}]$ without estimating the spherical PS  $P(k)$.  A similar approach was taken in \citetalias{AE23} (Section 6) in the context of foreground avoidance. The idea here is to use the full available information to find the maximum likelihood solution for the only parameter $[\Omega_{\HI}b_{\HI}]^2$. We have modelled $[C_{\ellb}(\Delta\nu)]_{\rm{res}}$ using equation~(\ref{eq:Cmodel}), and quantify  the $21$-cm signal $\left[C_{\ell}(\Delta\nu)\right]_{T}$ using equation~(\ref{eq:cl_Pk_sinc}). The formalism allows us to parameterize the entire $21$-cm signal with the parameter  $[\Omega_{\HI}b_{\HI}]^2$ which we estimate by the MLE. We have used $\left[C_{\ell}(\Delta\nu)\right]_{TC}$ instead of $\left[C_{\ell}(\Delta\nu)\right]_{T}$ as the model $21$-cm signal to account for the signal loss. 

\begin{table}
\centering
\caption{The upper limits on \omb{} from  PF.} 
    \begin{tabular}{ccccc}
        \hline
        \hline
        Set & \begin{tabular}[c]{@{}c@{}} $\ell$-range \end{tabular} &
        \begin{tabular}[c]{@{}c@{}} \ombsq{}\\$\times 10^{-4}$ \end{tabular} & SNR & \begin{tabular}[c]{@{}c@{}} \omb{}$_{\rm UL}$\\$\times 10^{-2}$ \end{tabular} \\
        \hline 
        I  & $<2000$ & $-3.13 \pm 2.53$ & $-1.24$ &   $2.249$ \\
        II  & $2000-4000$ & $6.33 \pm 4.49$ & $1.41$ &   $3.913$ \\
        III  & $>4000$ & $44.67 \pm 12.87$ & $3.47$ &   $8.391$ \\
        \rm{Combined}  & $<6300$ & $0.44 \pm 2.17$ & $0.20$ &   $\mathbf{2.187}$ \\

        \hline
    \end{tabular}
    \label{tab:results}
\end{table}

\begin{table}
\centering
\caption{The upper limits on \omb{} from GPR.} 
    \begin{tabular}{ccccc}
        \hline
        \hline
        Set & \begin{tabular}[c]{@{}c@{}} $\ell$-range \end{tabular} &
        \begin{tabular}[c]{@{}c@{}} \ombsq{}\\$\times 10^{-4}$ \end{tabular} & SNR & \begin{tabular}[c]{@{}c@{}} \omb{}$_{\rm UL}$\\$\times 10^{-2}$ \end{tabular} \\
        \hline
        I  & $<2000$ & $3.55 \pm 5.46$ & $0.65$ &   $3.805$ \\
        II  & $2000-4000$ & $-3.59 \pm 6.63$ & $-0.54$ &   $3.641$ \\
        III  & $>4000$ & $12.67 \pm 16.50$ & $0.77$ &   $6.758$ \\
        \rm{Combined}  & $<6300$ & $1.40 \pm 4.09$ & $0.34$ &   $\mathbf{3.093}$ \\
        \hline
    \end{tabular}
    \label{tab:results_GPR}
\end{table}

As mentioned in Section~\ref{sec:fitandres}, we have divided the measured \clb{} into three sets. We have applied the MLE on the $[C_{\ellb}(\Delta\nu)]_{\rm{res}}$ values from the individual sets  to constrain \ombsq{} for which the results for PF are presented in Table~\ref{tab:results}. Here  Set I yields a tight constraint  where we have  $[\Omega_{\HI}b_{\HI}]^2 = -3.13\times 10^{-4} \pm 2.53 \times 10^{-4}$ with the corresponding upper limit $[\Omega_{\HI}b_{\HI}]_{\rm UL} \leq 2.25 \times 10^{-2}$ at $2\sigma$ level. Note that while calculating the upper limit we have conservatively set $[\Omega_{\HI}b_{\HI}]^2$ to zero when it is negative. We  note that Set II also yields a considerably  tight upper limit $[\Omega_{\HI}b_{\HI}]_{\rm UL} \leq 3.91 \times 10^{-2}$ on the \HI{} abundance. Set III, however, gives a relatively poor constraint $[\Omega_{\HI}b_{\HI}]_{\rm UL} \leq 8.39 \times 10^{-2}$. This weak upper limit from Set III can be attributed to larger noise and residual foregrounds in these \clb{} values. We have also combined  $[C_{\ellb}(\Delta\nu)]_{\rm{res}}$ for all the three sets  to maximize the SNR and constrain $[\Omega_{\HI}b_{\HI}]^2$.  We find $[\Omega_{\HI}b_{\HI}]^2 = 4.4\times 10^{-5} \pm 2.17 \times 10^{-4}$, which yields the  tightest constraint $[\Omega_{\HI}b_{\HI}]_{\rm UL} \leq 2.19 \times 10^{-2}$ on the upper limit. The best-fit values obtained from the combined set are also presented in Table~\ref{tab:results}. 

The results from GPR are presented in Table~\ref{tab:results_GPR}. 
Here the constraints from Set II are better than those from Sets I and III, while the tightest limit from GPR  $[\Omega_{\HI}b_{\HI}]_{\rm UL} \leq 3.09 \times 10^{-2}$ is obtained by combining all three sets. Comparing Tables~\ref{tab:results} and ~\ref{tab:results_GPR},  we see that GPR  preforms better than PF for Sets II and III, whereas PF yields tighter constraints for Set I and when all the sets are combined.  Further,  we also find that the SNR values are always smaller for GPR for which all the $[\Omega_{\HI}b_{\HI}]^2$  are within  $0 \pm \sigma$.

\section{Summary and Conclusions}
\label{sec:conclusion}

This is the third in a series of papers which focus on $21$-cm IM considering  uGMRT data    centred at $432.8\,\rm{MHz}$ which corresponds to $z=2.28$. The data analyzed here is described in Section~\ref{sec:data}, and also our earlier work \citetalias{P22}.  In the present work we have used the MAPS \cl{} estimated in  \citetalias{AE23} using the `Cross' TGE which uses the cross-correlation of  the RR and LL  polarizations. This has the advantage of naturally avoiding  the noise bias and other systematics  which are uncorrelated between the two polarizations.  While both of our earlier works used foreground avoidance, the present work implements foreground removal which does away with the foreground wedge allowing us to use  the entire $(k_{\perp}, k_{\parallel})$ plane for estimating the $21$-cm PS.

The foregrounds generally  exhibit a smooth spectral  behaviour, and the contribution to the  measured \cl{} is expected to remain correlated even at large $\Delta \nu$.  In contrast, the predicted  $21$-cm signal $[C_{\ell}(\Delta\nu)]_T$  decorrelates rapidly with increasing  $\Delta\nu$,  and is close to zero beyond a characteristic frequency scale $\Delta \nu > [\Delta \nu]$. We find $[\Delta \nu] \sim 0.5  - 1.0 \, \rm{MHz}$  for most of  the $\ell$ range considered here (Figure~\ref{fig:predicted_signal}). 

Here we have considered two different approaches for foreground modelling and subtraction, the first being polynomial fitting (PF) and the second being Gaussian Process Regression (GPR). For both, we have used the range $\Delta\nu > [\Delta\nu]$  to estimate  the foreground contribution to  \clb{}.In PF  this is  modelled  as  an even polynomial in $\Delta \nu$. The values of $[\Delta \nu]$ and the polynomial order  are different for each $\ellb$, and the details are presented in Section~\ref{sec:fgmodelling}. For GPR, the details of foreground modelling  are  presented in Section~\ref{sec:fgmodelling_GPR}. For both PF and GPR, we have extrapolated the foreground model predictions to the range $\Delta\nu \le  [\Delta\nu]$, and subtracted out the foreground predictions from \clb. We have used the residual $[C_{\ellb}(\Delta \nu)]_{\rm res}$, after foreground subtraction, in the range  $\Delta\nu \le  [\Delta\nu]$ to constrain the $21$-cm signal. The foreground subtraction introduces a loss in the $21$-cm signal, which we have corrected for in the quantitative results presented in this work. The cylindrical PS $P(k_{\perp}, k_{\parallel})$, estimated from $[C_{\ellb}(\Delta \nu)]_{\rm res}$, appears to be largely free of foreground contribution, and  there is no indication of the foreground wedge (Figure~\ref{fig:cylps}). The noise statistics, quantified by $X$ (equation~\ref{eq:xstat}), are well described by a Lorentzian distribution which is largely symmetric around $X=0$. The distribution of $X$ does not exhibit any  large outliers, both positive or negative, indicating that the estimated PS is free of systematics.

The residual $[C_{\ellb}(\Delta \nu)]_{\rm res}$ has been used to estimate the spherical PS $P(k)$ using MLE which utilizes the statistical isotropy of the expected $21$-cm signal (Section~\ref{sec:BMLE}). The estimated $\Delta^2(k)$ exhibits values of both sign, however the negative values are within  $0\pm2\sigma$ and we  interpret these  as arising from statistical fluctuations.  We have found that the $\mid \Delta^2(k) \mid$ values and their uncertainties increase monotonically with increasing $k$. The  results from PF and GPR are  presented  in Tables~\ref{tab:ul_MLE} and ~\ref{tab:ul_GPR} respectively. Our best result $\Delta_{\rm UL}^2(k)\leq (18.07)^2 \,\rm{mK}^2$  comes from PF at the smallest $k$ bin $k = 0.247 \, \rm{Mpc}^{-1}$. This upper limit corresponds to $[\Omega_{\HI}b_{\HI}]_{\rm UL} \le  0.036$. Considering GPR, the corresponding values are  $\Delta_{\rm UL}^2(k)\leq (24.54)^2 \,\rm{mK}^2$ and   $[\Omega_{\HI}b_{\HI}]_{\rm UL} \le  0.045$ at $k = 0.296 \, \rm{Mpc}^{-1}$ which is the smallest $k$ bin for GPR.  Although the best upper limit comes from PF, we find that overall the $\Delta_{\rm UL}^2(k)$ values  from PF and GPR are quite comparable (Figure~\ref{fig:pssph}). 

In a different approach, we have combined  the estimated $[C_{\ellb}(\Delta \nu)]_{\rm res}$  to directly constrain the single parameter \omb{}  (Section~\ref{sec:omb}), for which the results  for PF and GPR are presented in Tables~\ref{tab:results} and ~\ref{tab:results_GPR} respectively. We see that PF provides the best upper limits.  Considering a particular subset of the $\ellb$ grids (Set I), we can place a tight upper limit $[\Omega_{\HI}b_{\HI}]_{\rm UL} \le  2.25 \times 10^{-2}$. It may however be noted  that this particular limit  incorporates a large correction factor to account for the $21$-cm signal loss due to foreground subtraction. The $\Delta \nu$ range used for estimating and extrapolating the foreground models is also different from that used for the other sets.  The tightest upper limit  $[\Omega_{\HI}b_{\HI}]_{\rm UL} \le  2.19 \times 10^{-2} $, however,  is obtained when we combine the entire $\ellb$ grid (Sets 1, II and  III).   This  limit is a factor of $3$ improvement  over the earlier works which have used  foreground avoidance. A comparison of the upper limits obtained from this uGMRT Band $3$ observation is presented  in Table~\ref{tab:results_IM}.

While we have removed the foregrounds to a large extent, we still have not detected the $21$-cm signal due to noise. On a positive note, we have put tight constraints on the upper limit of the $21$-cm IM signal.  Several different studies, both observational (e.g. \citealt{Rhee2018, AdityaC2020,ho2021, chime22, Cunnington23}) and theoretical (e.g. \citealt{hamsa2015,Deb16}) indicate that  $[\Omega_{\HI}b_{\HI}] \sim 10^{-3}$ for the redshift we have considered here. Our present upper limits on $[\Omega_{\HI} b_{\HI}]_{\rm UL} \leq 0.022$ is $\sim 10$ times larger than currently estimated values.  However, this upper limit is nearly $3$ times tighter over previous IM measurements at this redshift.

\section*{Acknowledgements}

We thank the anonymous reviewer for the detailed comments which helped us to improve the work. We thank the staff of GMRT for making this observation possible. GMRT is run by National Centre for Radio Astrophysics (NCRA) of the Tata Institute of Fundamental Research (TIFR). AG would like to thank IUCAA, Pune for providing support through the associateship programme. SB would like to acknowledge funding provided under the MATRICS grant SERB/F/9805/2019-2020 and AG would like to acknowledge funding provided under the SERB-SURE grant SUR/2022/000595 of the Science \& Engineering Research Board, a statutory body of Department of Science \& Technology (DST), Government of India. Part of this work has used the Supercomputing facility `PARAM Shakti' of IIT Kharagpur established under National Supercomputing Mission (NSM), Government of India and supported by Centre for Development of Advanced Computing (CDAC), Pune.

\section*{Data Availability}

The observed data are publicly available through the GMRT data archive\footnote{\url{https://naps.ncra.tifr.res.in/goa/}} under the proposal code $32\_120$. The simulated data used here are available upon reasonable request to the corresponding author.



\bibliographystyle{mnras}
\bibliography{myref} 

\begin{thebibliography}{}
\makeatletter
\relax
\def\mn@urlcharsother{\let\do\@makeother \do\$\do\&\do\#\do\^\do\_\do\%\do\~}
\def\mn@doi{\begingroup\mn@urlcharsother \@ifnextchar [ {\mn@doi@}
  {\mn@doi@[]}}
\def\mn@doi@[#1]#2{\def\@tempa{#1}\ifx\@tempa\@empty \href
  {http://dx.doi.org/#2} {doi:#2}\else \href {http://dx.doi.org/#2} {#1}\fi
  \endgroup}
\def\mn@eprint#1#2{\mn@eprint@#1:#2::\@nil}
\def\mn@eprint@arXiv#1{\href {http://arxiv.org/abs/#1} {{\tt arXiv:#1}}}
\def\mn@eprint@dblp#1{\href {http://dblp.uni-trier.de/rec/bibtex/#1.xml}
  {dblp:#1}}
\def\mn@eprint@#1:#2:#3:#4\@nil{\def\@tempa {#1}\def\@tempb {#2}\def\@tempc
  {#3}\ifx \@tempc \@empty \let \@tempc \@tempb \let \@tempb \@tempa \fi \ifx
  \@tempb \@empty \def\@tempb {arXiv}\fi \@ifundefined
  {mn@eprint@\@tempb}{\@tempb:\@tempc}{\expandafter \expandafter \csname
  mn@eprint@\@tempb\endcsname \expandafter{\@tempc}}}

\bibitem[\protect\citeauthoryear{Abdurashidova et~al.,}{Abdurashidova
  et~al.}{2022}]{Abdurashidova_2022}
Abdurashidova Z.,  et~al., 2022, \mn@doi [The Astrophysical Journal]
  {10.3847/1538-4357/ac1c78}, 925, 221

\bibitem[\protect\citeauthoryear{Ali, Bharadwaj  \& Chengalur}{Ali
  et~al.}{2008}]{ali}
Ali S.~S.,  Bharadwaj S.,   Chengalur J.~N.,  2008, \mn@doi [\mnras]
  {10.1111/j.1365-2966.2008.12984.x}, 385, 2166

\bibitem[\protect\citeauthoryear{{Ambikasaran}, {Foreman-Mackey}, {Greengard},
  {Hogg}  \& {O'Neil}}{{Ambikasaran} et~al.}{2015}]{george}
{Ambikasaran} S.,  {Foreman-Mackey} D.,  {Greengard} L.,  {Hogg} D.~W.,
  {O'Neil} M.,  2015, \mn@doi [IEEE Transactions on Pattern Analysis and
  Machine Intelligence] {10.1109/TPAMI.2015.2448083}, \href
  {https://ui.adsabs.harvard.edu/abs/2015ITPAM..38..252A} {38, 252}

\bibitem[\protect\citeauthoryear{{Amiri} et~al.,}{{Amiri}
  et~al.}{2023}]{chime22}
{Amiri} M.,  et~al., 2023, \mn@doi [\apj] {10.3847/1538-4357/acb13f}, \href
  {https://ui.adsabs.harvard.edu/abs/2023ApJ...947...16A} {947, 16}

\bibitem[\protect\citeauthoryear{{Ansari} et~al.,}{{Ansari}
  et~al.}{2012}]{ansari2012}
{Ansari} R.,  et~al., 2012, \mn@doi [\aap] {10.1051/0004-6361/201117837}, \href
  {https://ui.adsabs.harvard.edu/abs/2012A&A...540A.129A} {540, A129}

\bibitem[\protect\citeauthoryear{{Bagla}, {Khandai}  \& {Datta}}{{Bagla}
  et~al.}{2010}]{bagla2010}
{Bagla} J.~S.,  {Khandai} N.,   {Datta} K.~K.,  2010, \mn@doi [\mnras]
  {10.1111/j.1365-2966.2010.16933.x}, \href
  {https://ui.adsabs.harvard.edu/abs/2010MNRAS.407..567B} {407, 567}

\bibitem[\protect\citeauthoryear{{Battye}, {Browne}, {Dickinson}, {Heron},
  {Maffei}  \& {Pourtsidou}}{{Battye} et~al.}{2013}]{battye2013}
{Battye} R.~A.,  {Browne} I.~W.~A.,  {Dickinson} C.,  {Heron} G.,  {Maffei} B.,
    {Pourtsidou} A.,  2013, \mn@doi [\mnras] {10.1093/mnras/stt1082}, \href
  {https://ui.adsabs.harvard.edu/abs/2013MNRAS.434.1239B} {434, 1239}

\bibitem[\protect\citeauthoryear{{Bharadwaj} \& {Ali}}{{Bharadwaj} \&
  {Ali}}{2005}]{BA5}
{Bharadwaj} S.,  {Ali} S.~S.,  2005, \mn@doi [\mnras]
  {10.1111/j.1365-2966.2004.08604.x}, \href
  {http://adsabs.harvard.edu/abs/2005MNRAS.356.1519B} {356, 1519}

\bibitem[\protect\citeauthoryear{{Bharadwaj} \& {Pandey}}{{Bharadwaj} \&
  {Pandey}}{2003}]{Bharadwaj2003}
{Bharadwaj} S.,  {Pandey} S.~K.,  2003, \mn@doi [Journal of Astrophysics and
  Astronomy] {10.1007/BF03012189}, \href
  {https://ui.adsabs.harvard.edu/abs/2003JApA...24...23B} {24, 23}

\bibitem[\protect\citeauthoryear{{Bharadwaj} \& {Sethi}}{{Bharadwaj} \&
  {Sethi}}{2001}]{BS01}
{Bharadwaj} S.,  {Sethi} S.~K.,  2001, \mn@doi [\japa] {10.1007/BF02702273},
  \href {http://adsabs.harvard.edu/abs/2001JApA...22..293B} {22, 293}

\bibitem[\protect\citeauthoryear{{Bharadwaj} \& {Srikant}}{{Bharadwaj} \&
  {Srikant}}{2004}]{bh_sri2004}
{Bharadwaj} S.,  {Srikant} P.~S.,  2004, \mn@doi [\japa] {10.1007/BF02702289},
  \href {https://ui.adsabs.harvard.edu/abs/2004JApA...25...67B} {25, 67}

\bibitem[\protect\citeauthoryear{{Bharadwaj}, {Nath}  \& {Sethi}}{{Bharadwaj}
  et~al.}{2001}]{BNS}
{Bharadwaj} S.,  {Nath} B.~B.,   {Sethi} S.~K.,  2001, \mn@doi [\japa]
  {10.1007/BF02933588}, \href
  {https://ui.adsabs.harvard.edu/abs/2001JApA...22...21B} {22, 21}

\bibitem[\protect\citeauthoryear{Bharadwaj, Sethi  \& Saini}{Bharadwaj
  et~al.}{2009}]{Bh09}
Bharadwaj S.,  Sethi S.~K.,   Saini T.~D.,  2009, \mn@doi [Phys. Rev. D]
  {10.1103/PhysRevD.79.083538}, 79, 083538

\bibitem[\protect\citeauthoryear{Bharadwaj, Pal, Choudhuri  \& Dutta}{Bharadwaj
  et~al.}{2018}]{Bh18}
Bharadwaj S.,  Pal S.,  Choudhuri S.,   Dutta P.,  2018, \mn@doi [\mnras]
  {10.1093/mnras/sty3501}, 483, 5694

\bibitem[\protect\citeauthoryear{Bishop}{Bishop}{2006}]{bishop}
Bishop C.~M.,  [2006], Pattern recognition and machine learning.
New York : Springer, [2006] {\textcopyright}2006, \url
  {https://link.springer.com/book/9780387310732}

\bibitem[\protect\citeauthoryear{{Bull}, {Camera}, {Raccanelli}, {Blake},
  {Ferreira}, {Santos}  \& {Schwarz}}{{Bull} et~al.}{2015a}]{SKA15}
{Bull} P.,  {Camera} S.,  {Raccanelli} A.,  {Blake} C.,  {Ferreira} P.,
  {Santos} M.,   {Schwarz} D.~J.,  2015a, in AASKA14. p.~24 (\mn@eprint {arXiv}
  {1501.04088})

\bibitem[\protect\citeauthoryear{Bull, Ferreira, Patel  \& Santos}{Bull
  et~al.}{2015b}]{Bull15}
Bull P.,  Ferreira P.~G.,  Patel P.,   Santos M.~G.,  2015b, \mn@doi [\apj]
  {10.1088/0004-637x/803/1/21}, 803, 21

\bibitem[\protect\citeauthoryear{{CHIME Collaboration} et~al.,}{{CHIME
  Collaboration} et~al.}{2022}]{chimeIM}
{CHIME Collaboration} et~al., 2022, \mn@doi [\apjs] {10.3847/1538-4365/ac6fd9},
  \href {https://ui.adsabs.harvard.edu/abs/2022ApJS..261...29C} {261, 29}

\bibitem[\protect\citeauthoryear{Chakraborty et~al.,}{Chakraborty
  et~al.}{2019a}]{Cha1}
Chakraborty A.,  et~al., 2019a, \mn@doi [\mnras] {10.1093/mnras/stz1580}, 487,
  4102

\bibitem[\protect\citeauthoryear{Chakraborty et~al.,}{Chakraborty
  et~al.}{2019b}]{Cha2}
Chakraborty A.,  et~al., 2019b, \mn@doi [\mnras] {10.1093/mnras/stz2533}, 490,
  243

\bibitem[\protect\citeauthoryear{Chakraborty et~al.,}{Chakraborty
  et~al.}{2021}]{Ch21}
Chakraborty A.,  et~al., 2021, \mn@doi [\apjl] {10.3847/2041-8213/abd17a}, 907,
  L7

\bibitem[\protect\citeauthoryear{Chang, Pen, Peterson  \& McDonald}{Chang
  et~al.}{2008}]{Chang08}
Chang T.-C.,  Pen U.-L.,  Peterson J.~B.,   McDonald P.,  2008, \mn@doi [Phys.
  Rev. Lett.] {10.1103/PhysRevLett.100.091303}, 100, 091303

\bibitem[\protect\citeauthoryear{{Chang}, {Pen}, {Bandura}  \&
  {Peterson}}{{Chang} et~al.}{2010}]{chang10}
{Chang} T.-C.,  {Pen} U.-L.,  {Bandura} K.,   {Peterson} J.~B.,  2010, \mn@doi
  [\nat] {10.1038/nature09187}, \href
  {https://ui.adsabs.harvard.edu/abs/2010Natur.466..463C} {466, 463}

\bibitem[\protect\citeauthoryear{{Chapman} et~al.,}{{Chapman}
  et~al.}{2012}]{chapman12}
{Chapman} E.,  et~al., 2012, \mn@doi [\mnras]
  {10.1111/j.1365-2966.2012.21065.x}, \href
  {http://adsabs.harvard.edu/abs/2012MNRAS.423.2518C} {423, 2518}

\bibitem[\protect\citeauthoryear{Chatterjee, Bharadwaj, Choudhuri, Sethi  \&
  Patwa}{Chatterjee et~al.}{2022}]{Chatterjee2022}
Chatterjee S.,  Bharadwaj S.,  Choudhuri S.,  Sethi S.,   Patwa A.~K.,  2022,
  \mn@doi [\mnras] {10.1093/mnras/stac3576}, 519, 2410

\bibitem[\protect\citeauthoryear{Choudhuri, Bharadwaj, Ghosh  \& Ali}{Choudhuri
  et~al.}{2014}]{samir14}
Choudhuri S.,  Bharadwaj S.,  Ghosh A.,   Ali S.~S.,  2014, \mn@doi [\mnras]
  {10.1093/mnras/stu2027}, 445, 4351

\bibitem[\protect\citeauthoryear{Choudhuri, Bharadwaj, Roy, Ghosh  \&
  Ali}{Choudhuri et~al.}{2016a}]{samir16}
Choudhuri S.,  Bharadwaj S.,  Roy N.,  Ghosh A.,   Ali S.~S.,  2016a, \mn@doi
  [\mnras] {10.1093/mnras/stw607}, 459, 151

\bibitem[\protect\citeauthoryear{Choudhuri, Bharadwaj, Chatterjee, Ali, Roy  \&
  Ghosh}{Choudhuri et~al.}{2016b}]{samir17}
Choudhuri S.,  Bharadwaj S.,  Chatterjee S.,  Ali S.~S.,  Roy N.,   Ghosh A.,
  2016b, \mn@doi [\mnras] {10.1093/mnras/stw2254}, 463, 4093

\bibitem[\protect\citeauthoryear{Choudhuri, Bharadwaj, Ali, Roy, Intema  \&
  Ghosh}{Choudhuri et~al.}{2017}]{samir17a}
Choudhuri S.,  Bharadwaj S.,  Ali S.~S.,  Roy N.,  Intema H.~T.,   Ghosh A.,
  2017, \mn@doi [\mnras: Letters] {10.1093/mnrasl/slx066}, 470, L11

\bibitem[\protect\citeauthoryear{Choudhuri, Ghosh, Roy, Bharadwaj, Intema  \&
  Ali}{Choudhuri et~al.}{2020}]{samir20}
Choudhuri S.,  Ghosh A.,  Roy N.,  Bharadwaj S.,  Intema H.~T.,   Ali S.~S.,
  2020, \mn@doi [\mnras] {10.1093/mnras/staa762}, 494, 1936

\bibitem[\protect\citeauthoryear{{Chowdhury}, {Kanekar}, {Chengalur}, {Sethi}
  \& {Dwarakanath}}{{Chowdhury} et~al.}{2020}]{AdityaC2020}
{Chowdhury} A.,  {Kanekar} N.,  {Chengalur} J.~N.,  {Sethi} S.,   {Dwarakanath}
  K.~S.,  2020, \mn@doi [\nat] {10.1038/s41586-020-2794-7}, \href
  {https://ui.adsabs.harvard.edu/abs/2020Natur.586..369C} {586, 369}

\bibitem[\protect\citeauthoryear{{Cunnington} et~al.,}{{Cunnington}
  et~al.}{2023a}]{Cunnington23}
{Cunnington} S.,  et~al., 2023a, \mn@doi [\mnras] {10.1093/mnras/stac3060},
  \href {https://ui.adsabs.harvard.edu/abs/2023MNRAS.518.6262C} {518, 6262}

\bibitem[\protect\citeauthoryear{{Cunnington} et~al.,}{{Cunnington}
  et~al.}{2023b}]{CunningtonMeerKLASS2023}
{Cunnington} S.,  et~al., 2023b, \mn@doi [\mnras] {10.1093/mnras/stad1567},
  \href {https://ui.adsabs.harvard.edu/abs/2023MNRAS.523.2453C} {523, 2453}

\bibitem[\protect\citeauthoryear{Datta, Choudhury  \& Bharadwaj}{Datta
  et~al.}{2007}]{KD07}
Datta K.~K.,  Choudhury T.~R.,   Bharadwaj S.,  2007, \mn@doi [\mnras]
  {10.1111/j.1365-2966.2007.11747.x}, 378, 119

\bibitem[\protect\citeauthoryear{{Datta}, {Bowman}  \& {Carilli}}{{Datta}
  et~al.}{2010}]{adatta10}
{Datta} A.,  {Bowman} J.~D.,   {Carilli} C.~L.,  2010, \mn@doi [\apj]
  {10.1088/0004-637X/724/1/526}, \href
  {https://ui.adsabs.harvard.edu/abs/2010ApJ...724..526D} {724, 526}

\bibitem[\protect\citeauthoryear{{Dawson} et~al.,}{{Dawson}
  et~al.}{2016}]{Dawson_2016}
{Dawson} K.~S.,  et~al., 2016, \mn@doi [\aj] {10.3847/0004-6256/151/2/44},
  \href {https://ui.adsabs.harvard.edu/abs/2016AJ....151...44D} {151, 44}

\bibitem[\protect\citeauthoryear{{Di Matteo}, {Perna}, {Abel}  \& {Rees}}{{Di
  Matteo} et~al.}{2002}]{dmat1}
{Di Matteo} T.,  {Perna} R.,  {Abel} T.,   {Rees} M.~J.,  2002, \mn@doi [\apj]
  {10.1086/324293}, \href
  {https://ui.adsabs.harvard.edu/abs/2002ApJ...564..576D} {564, 576}

\bibitem[\protect\citeauthoryear{Dillon et~al.,}{Dillon
  et~al.}{2014}]{dillon14}
Dillon J.~S.,  et~al., 2014, \mn@doi [Phys. Rev. D]
  {10.1103/PhysRevD.89.023002}, 89, 023002

\bibitem[\protect\citeauthoryear{Dillon et~al.,}{Dillon
  et~al.}{2015}]{dillon15}
Dillon J.~S.,  et~al., 2015, \mn@doi [Phys. Rev. D]
  {10.1103/PhysRevD.91.123011}, 91, 123011

\bibitem[\protect\citeauthoryear{{Eisenstein} \& {Hu}}{{Eisenstein} \&
  {Hu}}{1998}]{Eisenstein_1998}
{Eisenstein} D.~J.,  {Hu} W.,  1998, \mn@doi [\apj] {10.1086/305424}, \href
  {https://ui.adsabs.harvard.edu/abs/1998ApJ...496..605E} {496, 605}

\bibitem[\protect\citeauthoryear{{Elahi} et~al.,}{{Elahi} et~al.}{2023}]{AE23}
{Elahi} K. M.~A.,  et~al., 2023, \mn@doi [\mnras] {10.1093/mnras/stad191},
  \href {https://ui.adsabs.harvard.edu/abs/2023MNRAS.520.2094E} {520, 2094}

\bibitem[\protect\citeauthoryear{{Ewall-Wice} et~al.,}{{Ewall-Wice}
  et~al.}{2021}]{Ewall-Wice2021}
{Ewall-Wice} A.,  et~al., 2021, \mn@doi [\mnras] {10.1093/mnras/staa3293},
  \href {https://ui.adsabs.harvard.edu/abs/2021MNRAS.500.5195E} {500, 5195}

\bibitem[\protect\citeauthoryear{{Gan} et~al.,}{{Gan} et~al.}{2022}]{Gan2022}
{Gan} H.,  et~al., 2022, \mn@doi [\aap] {10.1051/0004-6361/202142945}, \href
  {https://ui.adsabs.harvard.edu/abs/2022A&A...663A...9G} {663, A9}

\bibitem[\protect\citeauthoryear{Ghosh, Bharadwaj, Ali  \& Chengalur}{Ghosh
  et~al.}{2011a}]{ghosh1}
Ghosh A.,  Bharadwaj S.,  Ali S.~S.,   Chengalur J.~N.,  2011a, \mn@doi
  [\mnras] {10.1111/j.1365-2966.2010.17853.x}, 411, 2426

\bibitem[\protect\citeauthoryear{Ghosh, Bharadwaj, Ali  \& Chengalur}{Ghosh
  et~al.}{2011b}]{ghosh2}
Ghosh A.,  Bharadwaj S.,  Ali S.~S.,   Chengalur J.~N.,  2011b, \mn@doi
  [\mnras] {10.1111/j.1365-2966.2011.19649.x}, 418, 2584

\bibitem[\protect\citeauthoryear{Ghosh, Prasad, Bharadwaj, Ali  \&
  Chengalur}{Ghosh et~al.}{2012}]{ghosh3}
Ghosh A.,  Prasad J.,  Bharadwaj S.,  Ali S.~S.,   Chengalur J.~N.,  2012,
  \mn@doi [\mnras] {10.1111/j.1365-2966.2012.21889.x}, 426, 3295

\bibitem[\protect\citeauthoryear{{Gupta} et~al.,}{{Gupta} et~al.}{2017}]{uGMRT}
{Gupta} Y.,  et~al., 2017, Current Science, \href
  {https://ui.adsabs.harvard.edu/abs/2017CSci..113..707G} {113, 707}

\bibitem[\protect\citeauthoryear{{Hazra} \& {Guha Sarkar}}{{Hazra} \& {Guha
  Sarkar}}{2012}]{Hazra2012}
{Hazra} D.~K.,  {Guha Sarkar} T.,  2012, \mn@doi [\prl]
  {10.1103/PhysRevLett.109.121301}, \href
  {https://ui.adsabs.harvard.edu/abs/2012PhRvL.109l1301H} {109, 121301}

\bibitem[\protect\citeauthoryear{{Ho}, {Bird}  \& {Garnett}}{{Ho}
  et~al.}{2021}]{ho2021}
{Ho} M.-F.,  {Bird} S.,   {Garnett} R.,  2021, \mn@doi [\mnras]
  {10.1093/mnras/stab2169}, \href
  {https://ui.adsabs.harvard.edu/abs/2021MNRAS.507..704H} {507, 704}

\bibitem[\protect\citeauthoryear{{Kennedy} \& {Bull}}{{Kennedy} \&
  {Bull}}{2021}]{Kennedy21}
{Kennedy} F.,  {Bull} P.,  2021, \mn@doi [\mnras] {10.1093/mnras/stab1814},
  \href {https://ui.adsabs.harvard.edu/abs/2021MNRAS.506.2638K} {506, 2638}

\bibitem[\protect\citeauthoryear{{Kennedy}, {Bull}, {Wilensky}, {Burba}  \&
  {Choudhuri}}{{Kennedy} et~al.}{2023}]{Kennedy2023}
{Kennedy} F.,  {Bull} P.,  {Wilensky} M.~J.,  {Burba} J.,   {Choudhuri} S.,
  2023, \mn@doi [\apjs] {10.3847/1538-4365/acc324}, \href
  {https://ui.adsabs.harvard.edu/abs/2023ApJS..266...23K} {266, 23}

\bibitem[\protect\citeauthoryear{{Kern} \& {Liu}}{{Kern} \&
  {Liu}}{2021}]{Kern2021}
{Kern} N.~S.,  {Liu} A.,  2021, \mn@doi [\mnras] {10.1093/mnras/staa3736},
  \href {https://ui.adsabs.harvard.edu/abs/2021MNRAS.501.1463K} {501, 1463}

\bibitem[\protect\citeauthoryear{{Kumar}, {Dutta}  \& {Roy}}{{Kumar}
  et~al.}{2020}]{jais2020}
{Kumar} J.,  {Dutta} P.,   {Roy} N.,  2020, \mn@doi [\mnras]
  {10.1093/mnras/staa1371}, \href
  {https://ui.adsabs.harvard.edu/abs/2020MNRAS.495.3683K} {495, 3683}

\bibitem[\protect\citeauthoryear{{Kumar}, {Dutta}, {Choudhuri}  \&
  {Roy}}{{Kumar} et~al.}{2022}]{jais22}
{Kumar} J.,  {Dutta} P.,  {Choudhuri} S.,   {Roy} N.,  2022, \mn@doi [\mnras]
  {10.1093/mnras/stac499}, \href
  {https://ui.adsabs.harvard.edu/abs/2022MNRAS.512..186K} {512, 186}

\bibitem[\protect\citeauthoryear{{Lanzetta}, {Wolfe}  \& {Turnshek}}{{Lanzetta}
  et~al.}{1995}]{Lanzetta95}
{Lanzetta} K.~M.,  {Wolfe} A.~M.,   {Turnshek} D.~A.,  1995, \mn@doi [\apj]
  {10.1086/175286}, \href
  {https://ui.adsabs.harvard.edu/abs/1995ApJ...440..435L} {440, 435}

\bibitem[\protect\citeauthoryear{Liu \& Tegmark}{Liu \& Tegmark}{2012}]{liu12}
Liu A.,  Tegmark M.,  2012, \mn@doi [\mnras]
  {10.1111/j.1365-2966.2011.19989.x}, 419, 3491

\bibitem[\protect\citeauthoryear{Liu, Parsons  \& Trott}{Liu
  et~al.}{2014a}]{liu14a}
Liu A.,  Parsons A.~R.,   Trott C.~M.,  2014a, \mn@doi [Phys. Rev. D]
  {10.1103/PhysRevD.90.023018}, 90, 023018

\bibitem[\protect\citeauthoryear{Liu, Parsons  \& Trott}{Liu
  et~al.}{2014b}]{liu14b}
Liu A.,  Parsons A.~R.,   Trott C.~M.,  2014b, \mn@doi [Phys. Rev. D]
  {10.1103/PhysRevD.90.023019}, 90, 023019

\bibitem[\protect\citeauthoryear{Loeb \& Wyithe}{Loeb \& Wyithe}{2008}]{loeb08}
Loeb A.,  Wyithe J. S.~B.,  2008, \mn@doi [Phys. Rev. Lett.]
  {10.1103/PhysRevLett.100.161301}, 100, 161301

\bibitem[\protect\citeauthoryear{{Long}, {Morales-Guti{\'e}rrez},
  {Montero-Camacho}  \& {Hirata}}{{Long} et~al.}{2022}]{long2022}
{Long} H.,  {Morales-Guti{\'e}rrez} C.,  {Montero-Camacho} P.,   {Hirata}
  C.~M.,  2022, arXiv e-prints, \href
  {https://ui.adsabs.harvard.edu/abs/2022arXiv221002385L} {p. arXiv:2210.02385}

\bibitem[\protect\citeauthoryear{Mazumder, Chakraborty, Datta, Choudhuri, Roy,
  Wadadekar  \& Ishwara-Chandra}{Mazumder et~al.}{2020}]{Mazumder20}
Mazumder A.,  Chakraborty A.,  Datta A.,  Choudhuri S.,  Roy N.,  Wadadekar Y.,
    Ishwara-Chandra C.~H.,  2020, \mn@doi [\mnras] {10.1093/mnras/staa1317},
  495, 4071

\bibitem[\protect\citeauthoryear{{McMullin}, {Waters}, {Schiebel}, {Young}  \&
  {Golap}}{{McMullin} et~al.}{2007}]{casa07}
{McMullin} J.~P.,  {Waters} B.,  {Schiebel} D.,  {Young} W.,   {Golap} K.,
  2007, in {Shaw} R.~A.,  {Hill} F.,   {Bell} D.~J.,  eds,  Astronomical
  Society of the Pacific Conference Series Vol. 376, Astronomical Data Analysis
  Software and Systems XVI. p.~127

\bibitem[\protect\citeauthoryear{{Mertens}, {Ghosh}  \& {Koopmans}}{{Mertens}
  et~al.}{2018}]{mertens18}
{Mertens} F.~G.,  {Ghosh} A.,   {Koopmans} L.~V.~E.,  2018, \mn@doi [\mnras]
  {10.1093/mnras/sty1207}, \href
  {https://ui.adsabs.harvard.edu/abs/2018MNRAS.478.3640M} {478, 3640}

\bibitem[\protect\citeauthoryear{Mertens et~al.,}{Mertens
  et~al.}{2020}]{mertens20}
Mertens F.~G.,  et~al., 2020, \mn@doi [\mnras] {10.1093/mnras/staa327}, 493,
  1662

\bibitem[\protect\citeauthoryear{{Mondal}, {Bharadwaj}  \& {Datta}}{{Mondal}
  et~al.}{2018}]{Mondal2018}
{Mondal} R.,  {Bharadwaj} S.,   {Datta} K.~K.,  2018, \mn@doi [\mnras]
  {10.1093/mnras/stx2888}, \href
  {https://ui.adsabs.harvard.edu/abs/2018MNRAS.474.1390M} {474, 1390}

\bibitem[\protect\citeauthoryear{{Mondal}, {Bharadwaj}, {Iliev}, {Datta},
  {Majumdar}, {Shaw}  \& {Sarkar}}{{Mondal} et~al.}{2019}]{Mondal19}
{Mondal} R.,  {Bharadwaj} S.,  {Iliev} I.~T.,  {Datta} K.~K.,  {Majumdar} S.,
  {Shaw} A.~K.,   {Sarkar} A.~K.,  2019, \mn@doi [\mnras]
  {10.1093/mnrasl/sly226}, \href
  {https://ui.adsabs.harvard.edu/abs/2019MNRAS.483L.109M} {483, L109}

\bibitem[\protect\citeauthoryear{Morales, Hazelton, Sullivan  \&
  Beardsley}{Morales et~al.}{2012}]{Morales_2012}
Morales M.~F.,  Hazelton B.,  Sullivan I.,   Beardsley A.,  2012, \mn@doi
  [\apj] {10.1088/0004-637x/752/2/137}, 752, 137

\bibitem[\protect\citeauthoryear{{Newburgh} et~al.,}{{Newburgh}
  et~al.}{2016}]{Newburgh16}
{Newburgh} L.~B.,  et~al., 2016, in {Hall} H.~J.,  {Gilmozzi} R.,   {Marshall}
  H.~K.,  eds,  Society of Photo-Optical Instrumentation Engineers (SPIE)
  Conference Series Vol. 9906, Ground-based and Airborne Telescopes VI. p.
  99065X (\mn@eprint {arXiv} {1607.02059}), \mn@doi{10.1117/12.2234286}

\bibitem[\protect\citeauthoryear{{Noterdaeme} et~al.,}{{Noterdaeme}
  et~al.}{2012}]{Not}
{Noterdaeme} P.,  et~al., 2012, \mn@doi [\aap] {10.1051/0004-6361/201220259},
  \href {https://ui.adsabs.harvard.edu/abs/2012A&A...547L...1N} {547, L1}

\bibitem[\protect\citeauthoryear{{Offringa}, {de Bruyn}, {Biehl}, {Zaroubi},
  {Bernardi}  \& {Pandey}}{{Offringa} et~al.}{2010}]{Off10}
{Offringa} A.~R.,  {de Bruyn} A.~G.,  {Biehl} M.,  {Zaroubi} S.,  {Bernardi}
  G.,   {Pandey} V.~N.,  2010, \mn@doi [\mnras]
  {10.1111/j.1365-2966.2010.16471.x}, \href
  {https://ui.adsabs.harvard.edu/abs/2010MNRAS.405..155O} {405, 155}

\bibitem[\protect\citeauthoryear{{Offringa}, {van de Gronde}  \&
  {Roerdink}}{{Offringa} et~al.}{2012}]{Off12}
{Offringa} A.~R.,  {van de Gronde} J.~J.,   {Roerdink} J.~B.~T.~M.,  2012,
  \mn@doi [\aap] {10.1051/0004-6361/201118497}, \href
  {https://ui.adsabs.harvard.edu/abs/2012A&A...539A..95O} {539, A95}

\bibitem[\protect\citeauthoryear{Paciga et~al.,}{Paciga
  et~al.}{2011}]{paciga11}
Paciga G.,  et~al., 2011, \mn@doi [\mnras] {10.1111/j.1365-2966.2011.18208.x},
  413, 1174

\bibitem[\protect\citeauthoryear{{Padmanabhan}, {Choudhury}  \&
  {Refregier}}{{Padmanabhan} et~al.}{2015}]{hamsa2015}
{Padmanabhan} H.,  {Choudhury} T.~R.,   {Refregier} A.,  2015, \mn@doi [\mnras]
  {10.1093/mnras/stu2702}, \href
  {https://ui.adsabs.harvard.edu/abs/2015MNRAS.447.3745P} {447, 3745}

\bibitem[\protect\citeauthoryear{{Pal}, {Bharadwaj}, {Ghosh}  \&
  {Choudhuri}}{{Pal} et~al.}{2021}]{Pal20}
{Pal} S.,  {Bharadwaj} S.,  {Ghosh} A.,   {Choudhuri} S.,  2021, \mn@doi
  [\mnras] {10.1093/mnras/staa3831}, \href
  {https://ui.adsabs.harvard.edu/abs/2021MNRAS.501.3378P} {501, 3378}

\bibitem[\protect\citeauthoryear{{Pal} et~al.,}{{Pal} et~al.}{2022}]{P22}
{Pal} S.,  et~al., 2022, \mn@doi [\mnras] {10.1093/mnras/stac2419}, \href
  {https://ui.adsabs.harvard.edu/abs/2022MNRAS.516.2851P} {516, 2851}

\bibitem[\protect\citeauthoryear{{Parsons} \& {Backer}}{{Parsons} \&
  {Backer}}{2009}]{Parsons_2009}
{Parsons} A.~R.,  {Backer} D.~C.,  2009, \mn@doi [\aj]
  {10.1088/0004-6256/138/1/219}, \href
  {https://ui.adsabs.harvard.edu/abs/2009AJ....138..219P} {138, 219}

\bibitem[\protect\citeauthoryear{{Paul}, {Santos}, {Chen}  \& {Wolz}}{{Paul}
  et~al.}{2023}]{Paul23}
{Paul} S.,  {Santos} M.~G.,  {Chen} Z.,   {Wolz} L.,  2023, \mn@doi [arXiv
  e-prints] {10.48550/arXiv.2301.11943}, \href
  {https://ui.adsabs.harvard.edu/abs/2023arXiv230111943P} {p. arXiv:2301.11943}

\bibitem[\protect\citeauthoryear{{Planck Collaboration} et~al.,}{{Planck
  Collaboration} et~al.}{2020}]{Planck18f}
{Planck Collaboration} et~al., 2020, \mn@doi [\aap]
  {10.1051/0004-6361/201833910}, \href
  {https://ui.adsabs.harvard.edu/abs/2020A&A...641A...6P} {641, A6}

\bibitem[\protect\citeauthoryear{{Pober} et~al.,}{{Pober}
  et~al.}{2013}]{pober13}
{Pober} J.~C.,  et~al., 2013, \mn@doi [\apjl] {10.1088/2041-8205/768/2/L36},
  \href {https://ui.adsabs.harvard.edu/abs/2013ApJ...768L..36P} {768, L36}

\bibitem[\protect\citeauthoryear{{Pober} et~al.,}{{Pober}
  et~al.}{2014}]{pober14}
{Pober} J.~C.,  et~al., 2014, \mn@doi [\apj] {10.1088/0004-637X/782/2/66},
  \href {https://ui.adsabs.harvard.edu/abs/2014ApJ...782...66P} {782, 66}

\bibitem[\protect\citeauthoryear{{Pober} et~al.,}{{Pober}
  et~al.}{2016}]{pober16}
{Pober} J.~C.,  et~al., 2016, \mn@doi [\apj] {10.3847/0004-637X/819/1/8}, \href
  {https://ui.adsabs.harvard.edu/abs/2016ApJ...819....8P} {819, 8}

\bibitem[\protect\citeauthoryear{Rasmussen \& Williams}{Rasmussen \&
  Williams}{2006}]{RW}
Rasmussen C.,  Williams C.,  2006, Gaussian Processes for Machine Learning.
Adaptive Computation and Machine Learning, MIT Press, Cambridge, MA, USA

\bibitem[\protect\citeauthoryear{{Rhee}, {Lah}, {Briggs}, {Chengalur},
  {Colless}, {Willner}, {Ashby}  \& {Le F{\`e}vre}}{{Rhee}
  et~al.}{2018}]{Rhee2018}
{Rhee} J.,  {Lah} P.,  {Briggs} F.~H.,  {Chengalur} J.~N.,  {Colless} M.,
  {Willner} S.~P.,  {Ashby} M. L.~N.,   {Le F{\`e}vre} O.,  2018, \mn@doi
  [\mnras] {10.1093/mnras/stx2461}, \href
  {https://ui.adsabs.harvard.edu/abs/2018MNRAS.473.1879R} {473, 1879}

\bibitem[\protect\citeauthoryear{{Saha}, {Bharadwaj}, {Roy}, {Choudhuri}  \&
  {Chattopadhyay}}{{Saha} et~al.}{2019}]{Preetha19}
{Saha} P.,  {Bharadwaj} S.,  {Roy} N.,  {Choudhuri} S.,   {Chattopadhyay} D.,
  2019, \mn@doi [\mnras] {10.1093/mnras/stz2528}, \href
  {https://ui.adsabs.harvard.edu/abs/2019MNRAS.489.5866S} {489, 5866}

\bibitem[\protect\citeauthoryear{{Saha}, {Bharadwaj}, {Chakravorty}, {Roy},
  {Choudhuri}, {G{\"u}nther}  \& {Smith}}{{Saha} et~al.}{2021}]{Preetha21}
{Saha} P.,  {Bharadwaj} S.,  {Chakravorty} S.,  {Roy} N.,  {Choudhuri} S.,
  {G{\"u}nther} H.~M.,   {Smith} R.~K.,  2021, \mn@doi [\mnras]
  {10.1093/mnras/stab446}, \href
  {https://ui.adsabs.harvard.edu/abs/2021MNRAS.502.5313S} {502, 5313}

\bibitem[\protect\citeauthoryear{{Saiyad Ali}, {Bharadwaj}  \&
  {Pandey}}{{Saiyad Ali} et~al.}{2006}]{Ali2006}
{Saiyad Ali} S.,  {Bharadwaj} S.,   {Pandey} S.~K.,  2006, \mn@doi [\mnras]
  {10.1111/j.1365-2966.2005.09847.x}, \href
  {https://ui.adsabs.harvard.edu/abs/2006MNRAS.366..213S} {366, 213}

\bibitem[\protect\citeauthoryear{{Santos}, {Cooray}  \& {Knox}}{{Santos}
  et~al.}{2005}]{santos05}
{Santos} M.~G.,  {Cooray} A.,   {Knox} L.,  2005, \mn@doi [\apj]
  {10.1086/429857}, \href
  {https://ui.adsabs.harvard.edu/abs/2005ApJ...625..575S} {625, 575}

\bibitem[\protect\citeauthoryear{Sarkar, Bharadwaj  \& Anathpindika}{Sarkar
  et~al.}{2016}]{Deb16}
Sarkar D.,  Bharadwaj S.,   Anathpindika S.,  2016, \mn@doi [\mnras]
  {10.1093/mnras/stw1111}, 460, 4310

\bibitem[\protect\citeauthoryear{{Shaver}, {Windhorst}, {Madau}  \& {de
  Bruyn}}{{Shaver} et~al.}{1999}]{shaver99}
{Shaver} P.~A.,  {Windhorst} R.~A.,  {Madau} P.,   {de Bruyn} A.~G.,  1999,
  \aap, \href {https://ui.adsabs.harvard.edu/abs/1999A&A...345..380S} {345,
  380}

\bibitem[\protect\citeauthoryear{{Slosar} et~al.,}{{Slosar}
  et~al.}{2019}]{puma}
{Slosar} A.,  et~al., 2019, in Bulletin of the American Astronomical Society.
  p.~53 (\mn@eprint {arXiv} {1907.12559})

\bibitem[\protect\citeauthoryear{{Subrahmanya}, {Manoharan}  \&
  {Chengalur}}{{Subrahmanya} et~al.}{2017}]{OWFA}
{Subrahmanya} C.~R.,  {Manoharan} P.~K.,   {Chengalur} J.~N.,  2017, \mn@doi
  [\japa] {10.1007/s12036-017-9430-4}, \href
  {https://ui.adsabs.harvard.edu/abs/2017JApA...38...10S} {38, 10}

\bibitem[\protect\citeauthoryear{{Swarup}, {Ananthakrishnan}, {Kapahi}, {Rao},
  {Subrahmanya}  \& {Kulkarni}}{{Swarup} et~al.}{1991}]{swarup91}
{Swarup} G.,  {Ananthakrishnan} S.,  {Kapahi} V.~K.,  {Rao} A.~P.,
  {Subrahmanya} C.~R.,   {Kulkarni} V.~K.,  1991, Current Science, \href
  {https://ui.adsabs.harvard.edu/abs/1991CuSc...60...95S} {60, 95}

\bibitem[\protect\citeauthoryear{Switzer et~al.,}{Switzer et~al.}{2013}]{SW13}
Switzer E.~R.,  et~al., 2013, \mn@doi [\mnras: Letters]
  {10.1093/mnrasl/slt074}, 434, L46

\bibitem[\protect\citeauthoryear{{Trott}}{{Trott}}{2016}]{Trott2016}
{Trott} C.~M.,  2016, \mn@doi [\mnras] {10.1093/mnras/stw1310}, \href
  {https://ui.adsabs.harvard.edu/abs/2016MNRAS.461..126T} {461, 126}

\bibitem[\protect\citeauthoryear{{Trott}, {Wayth}  \& {Tingay}}{{Trott}
  et~al.}{2012}]{trott1}
{Trott} C.~M.,  {Wayth} R.~B.,   {Tingay} S.~J.,  2012, \mn@doi [\apj]
  {10.1088/0004-637X/757/1/101}, \href
  {https://ui.adsabs.harvard.edu/abs/2012ApJ...757..101T} {757, 101}

\bibitem[\protect\citeauthoryear{{Trott}, {Mondal}, {Mellema}, {Murray},
  {Greig}, {Line}, {Barry}  \& {Morales}}{{Trott} et~al.}{2022}]{Trott2022}
{Trott} C.~M.,  {Mondal} R.,  {Mellema} G.,  {Murray} S.~G.,  {Greig} B.,
  {Line} J. L.~B.,  {Barry} N.,   {Morales} M.~F.,  2022, \mn@doi [\aap]
  {10.1051/0004-6361/202244024}, \href
  {https://ui.adsabs.harvard.edu/abs/2022A&A...666A.106T} {666, A106}

\bibitem[\protect\citeauthoryear{{Vedantham}, {Udaya Shankar}  \&
  {Subrahmanyan}}{{Vedantham} et~al.}{2012}]{vedantham12}
{Vedantham} H.,  {Udaya Shankar} N.,   {Subrahmanyan} R.,  2012, \mn@doi [\apj]
  {10.1088/0004-637X/745/2/176}, \href
  {https://ui.adsabs.harvard.edu/abs/2012ApJ...745..176V} {745, 176}

\bibitem[\protect\citeauthoryear{Visbal, Loeb  \& Wyithe}{Visbal
  et~al.}{2009}]{Visbal_2009}
Visbal E.,  Loeb A.,   Wyithe S.,  2009, \mn@doi [\jcap]
  {10.1088/1475-7516/2009/10/030}, 2009, 030

\bibitem[\protect\citeauthoryear{{Wilensky}, {Brown}  \& {Hazelton}}{{Wilensky}
  et~al.}{2023}]{Wilensky23}
{Wilensky} M.~J.,  {Brown} J.,   {Hazelton} B.~J.,  2023, \mn@doi [\mnras]
  {10.1093/mnras/stad863}, \href
  {https://ui.adsabs.harvard.edu/abs/2023MNRAS.521.5191W} {521, 5191}

\bibitem[\protect\citeauthoryear{Williams \& Rasmussen}{Williams \&
  Rasmussen}{1996}]{WR1995}
Williams C.,  Rasmussen C.,  1996, in Advances in neural information processing
  systems 8. MIT Press, Cambridge, MA, USA, pp 514--520

\bibitem[\protect\citeauthoryear{{Wolfe}, {Lanzetta}, {Foltz}  \&
  {Chaffee}}{{Wolfe} et~al.}{1995}]{Wolfe95}
{Wolfe} A.~M.,  {Lanzetta} K.~M.,  {Foltz} C.~B.,   {Chaffee} F.~H.,  1995,
  \mn@doi [\apj] {10.1086/176523}, \href
  {https://ui.adsabs.harvard.edu/abs/1995ApJ...454..698W} {454, 698}

\bibitem[\protect\citeauthoryear{Wolz et~al.,}{Wolz et~al.}{2021}]{Wolz2021}
Wolz L.,  et~al., 2021, \mn@doi [\mnras] {10.1093/mnras/stab3621}, 510, 3495

\bibitem[\protect\citeauthoryear{Wyithe, Loeb  \& Geil}{Wyithe
  et~al.}{2008}]{w08}
Wyithe J. S.~B.,  Loeb A.,   Geil P.~M.,  2008, \mn@doi [\mnras]
  {10.1111/j.1365-2966.2007.12631.x}, 383, 1195

\bibitem[\protect\citeauthoryear{{Zafar}, {P{\'e}roux}, {Popping}, {Milliard},
  {Deharveng}  \& {Frank}}{{Zafar} et~al.}{2013}]{zafar}
{Zafar} T.,  {P{\'e}roux} C.,  {Popping} A.,  {Milliard} B.,  {Deharveng}
  J.~M.,   {Frank} S.,  2013, \mn@doi [\aap] {10.1051/0004-6361/201321154},
  \href {https://ui.adsabs.harvard.edu/abs/2013A&A...556A.141Z} {556, A141}

\bibitem[\protect\citeauthoryear{{Zaldarriaga}, {Furlanetto}  \&
  {Hernquist}}{{Zaldarriaga} et~al.}{2004}]{Zaldarriaga2004}
{Zaldarriaga} M.,  {Furlanetto} S.~R.,   {Hernquist} L.,  2004, \mn@doi [\apj]
  {10.1086/386327}, \href
  {https://ui.adsabs.harvard.edu/abs/2004ApJ...608..622Z} {608, 622}

\makeatother
\end{thebibliography}



\appendix

\section{Results: Set I, II and III}
\label{sec:sec2n3}

\begin{figure}
    \centering
    \includegraphics[width=\columnwidth]{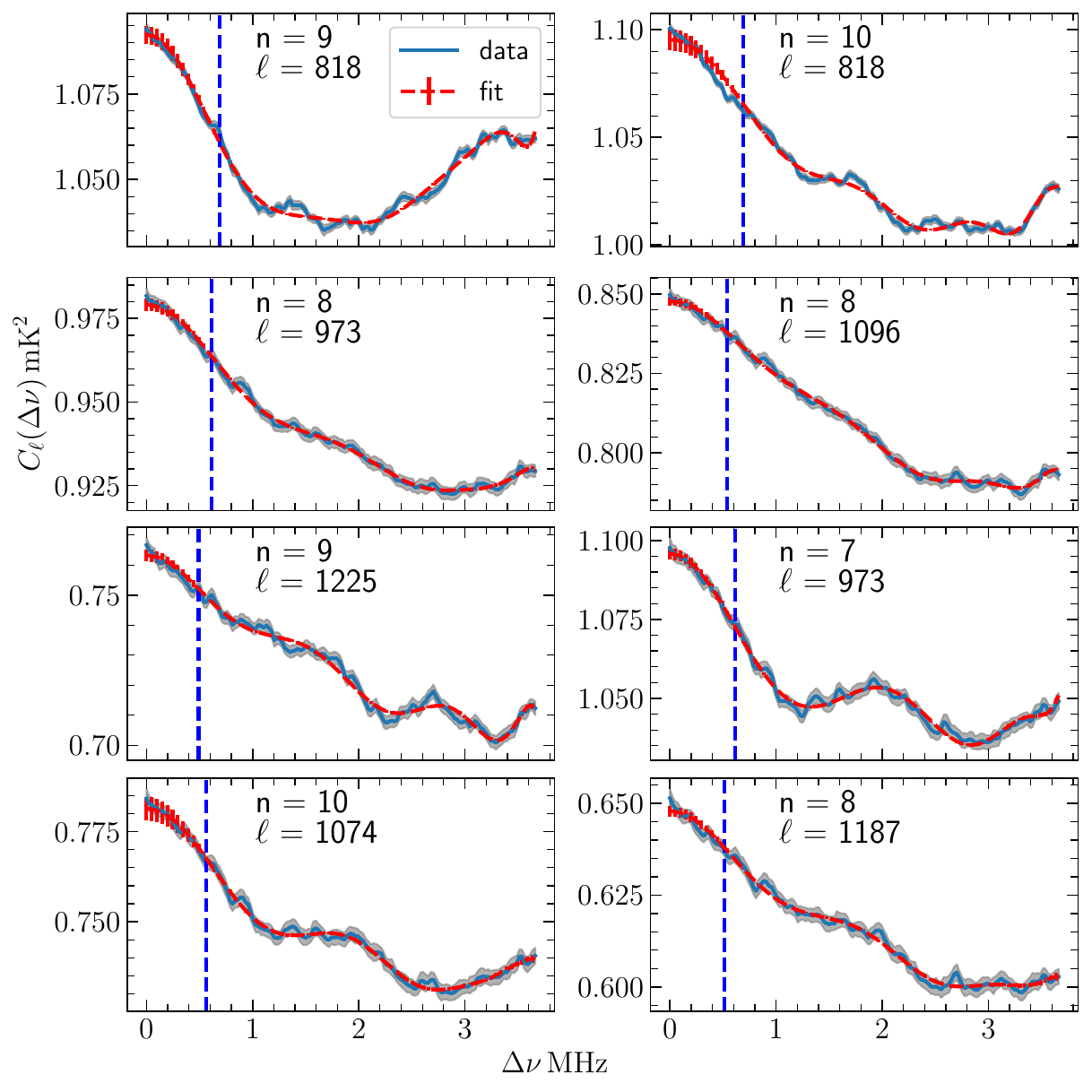}
    \caption{For Set I, the blue solid lines  show the measured \clb{} with $2\sigma$ uncertainties (grey shaded region) expected from scaled system noise. The red dashed lines and the associated error bars show the best-fit foreground models for PF and their $2\sigma$ uncertainties. The value of $\ellb$ and $n$ (used for foreground modelling) are mentioned in the respective panels. The vertical blue dashed lines show $[\Delta\nu]_{0.4}$.} 
    \label{fig:fg_fit_s1a}
\end{figure}

\begin{figure}
    \centering
    \includegraphics[width=\columnwidth]{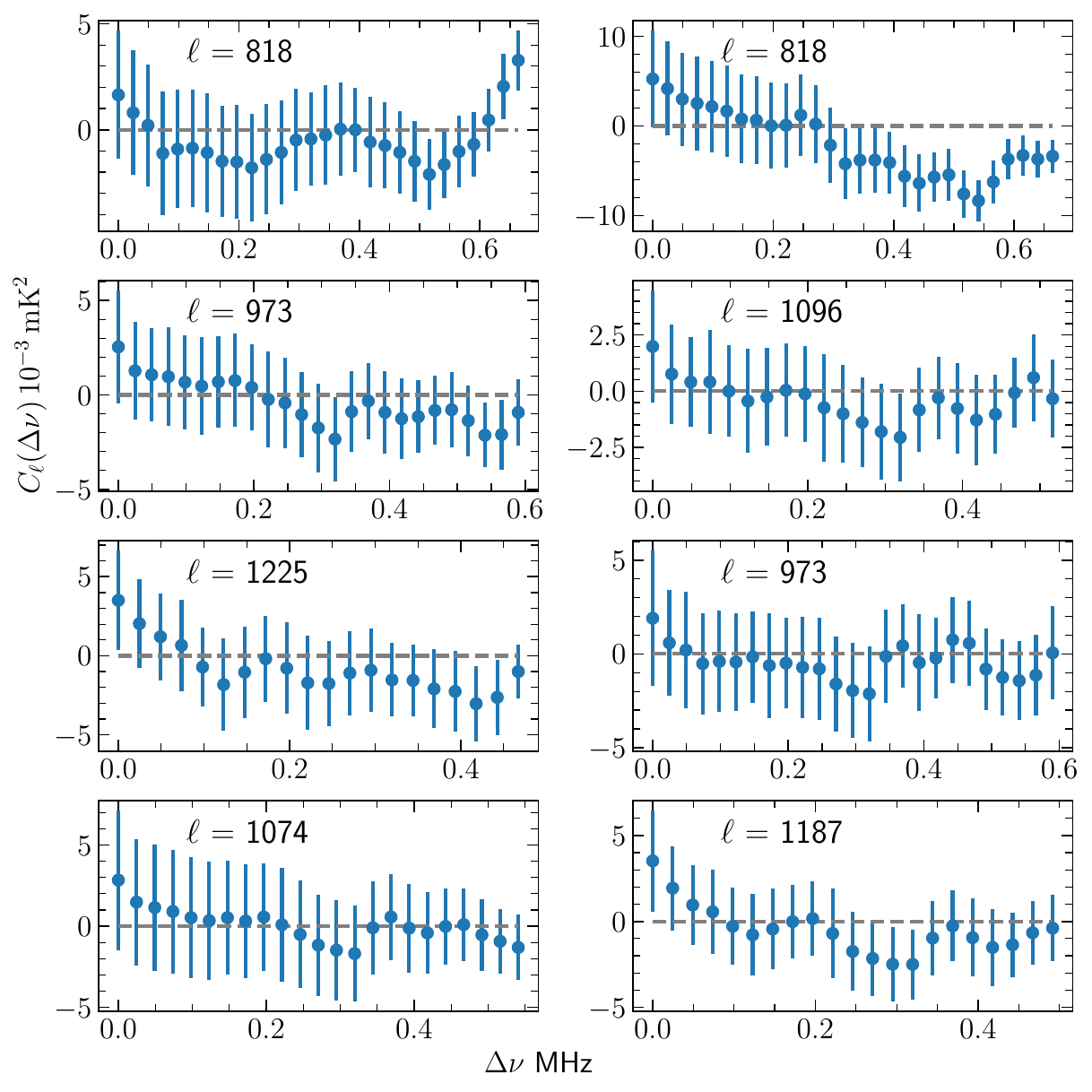}
    \caption{For Set I, the residual $\left[C_{\ellb}(\Delta\nu)\right]_{\rm{res}}$ in the range $\Delta\nu \le [\Delta\nu]_{0.4}$ with the $2\sigma$ error bars (combining  the  noise and fitting  errors), for the $\ellb$ quoted in the panel. }
    \label{fig:model_s1a}
\end{figure}

\begin{figure}
    \centering
    \includegraphics[width=\columnwidth]{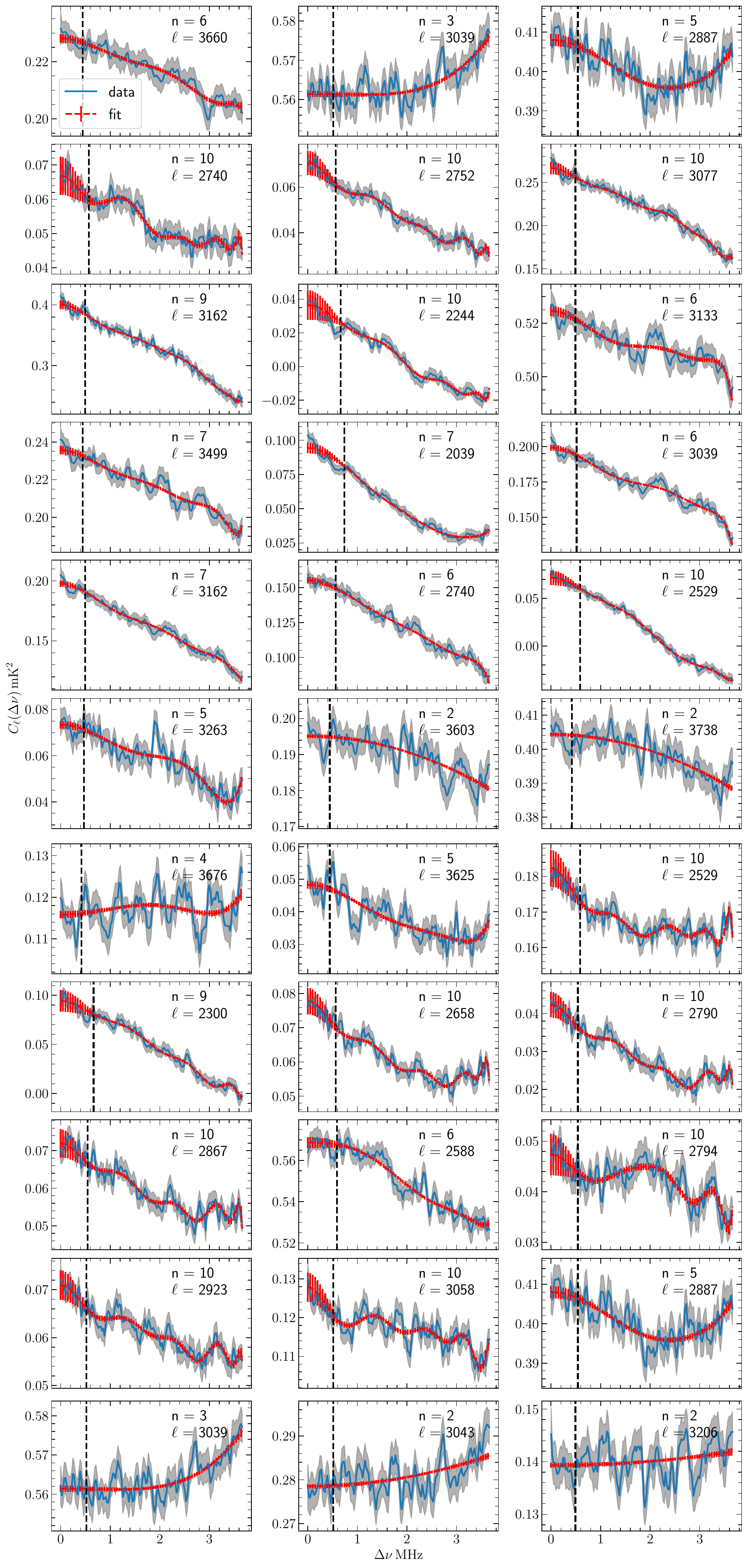}
    \caption{Same as Figure~\ref{fig:fg_fit_s1a} but for Set II. Here the vertical black dashed line shows $[\Delta\nu]_{0.1}$.}
    \label{fig:fg_fit_s2}
\end{figure}

\begin{figure}
    \centering
    \includegraphics[width=\columnwidth]{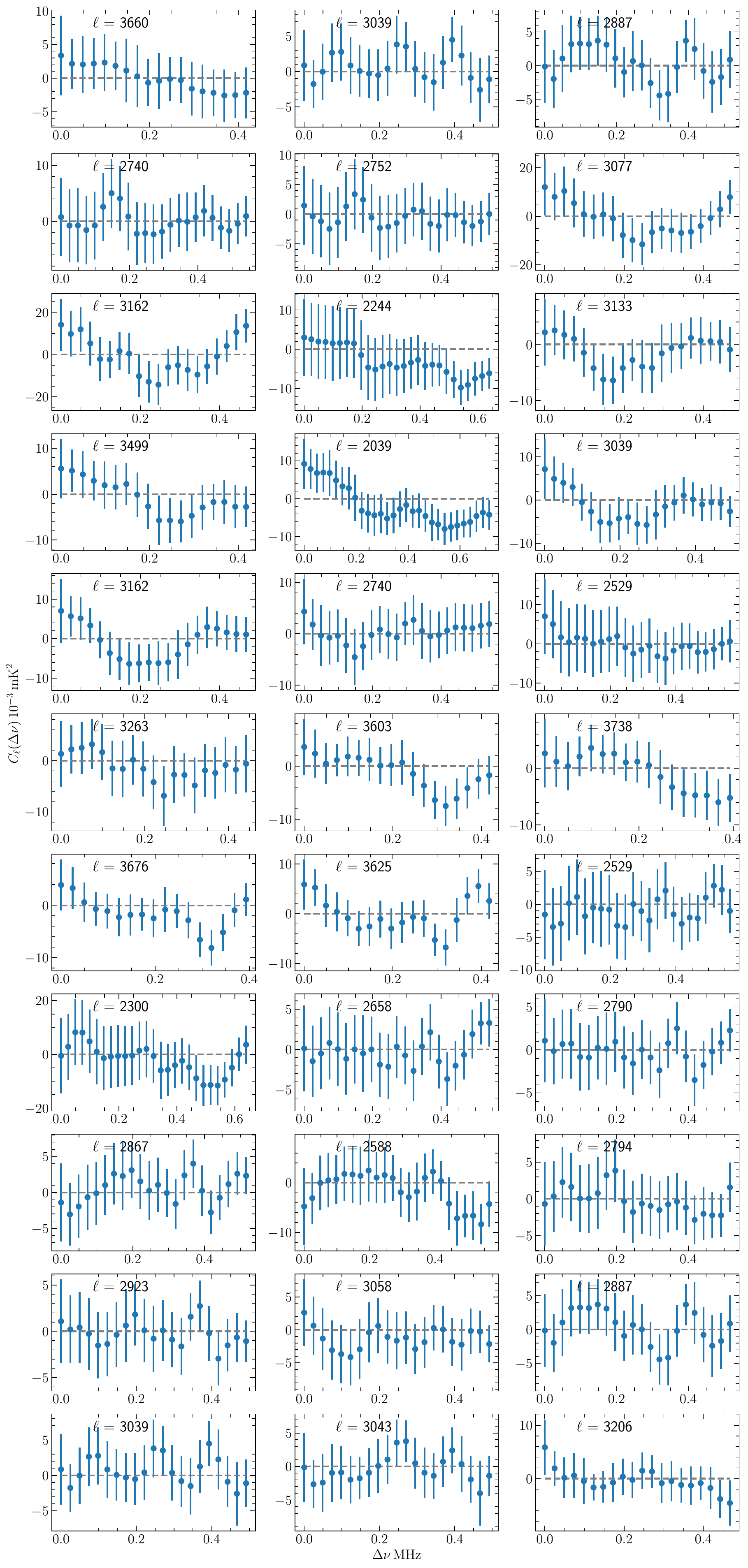}
    \caption{Same as Figure~\ref{fig:model_s1a}  but for Set II considering  the range $\Delta\nu \le  [\Delta\nu]_{0.1}$}
    \label{fig:model_s2}
\end{figure}

\begin{figure}
    \centering
    \includegraphics[width=\columnwidth]{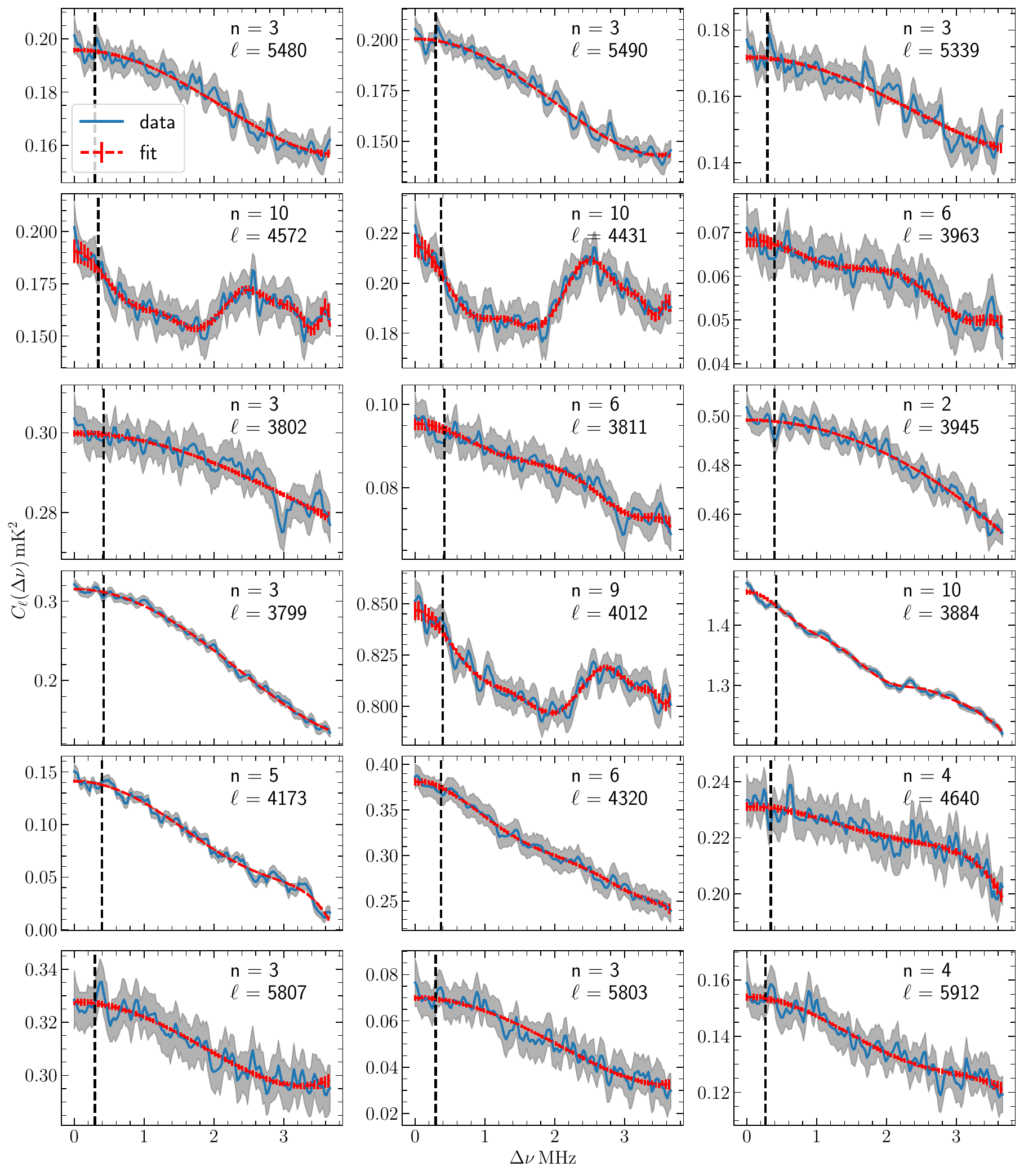}
    \caption{Same as Figure~\ref{fig:fg_fit_s1a} but for Set III. Here the vertical black dashed line shows $[\Delta\nu]_{0.1}$.}
    \label{fig:fg_fit_s3}
\end{figure}

\begin{figure}
    \centering
    \includegraphics[width=\columnwidth]{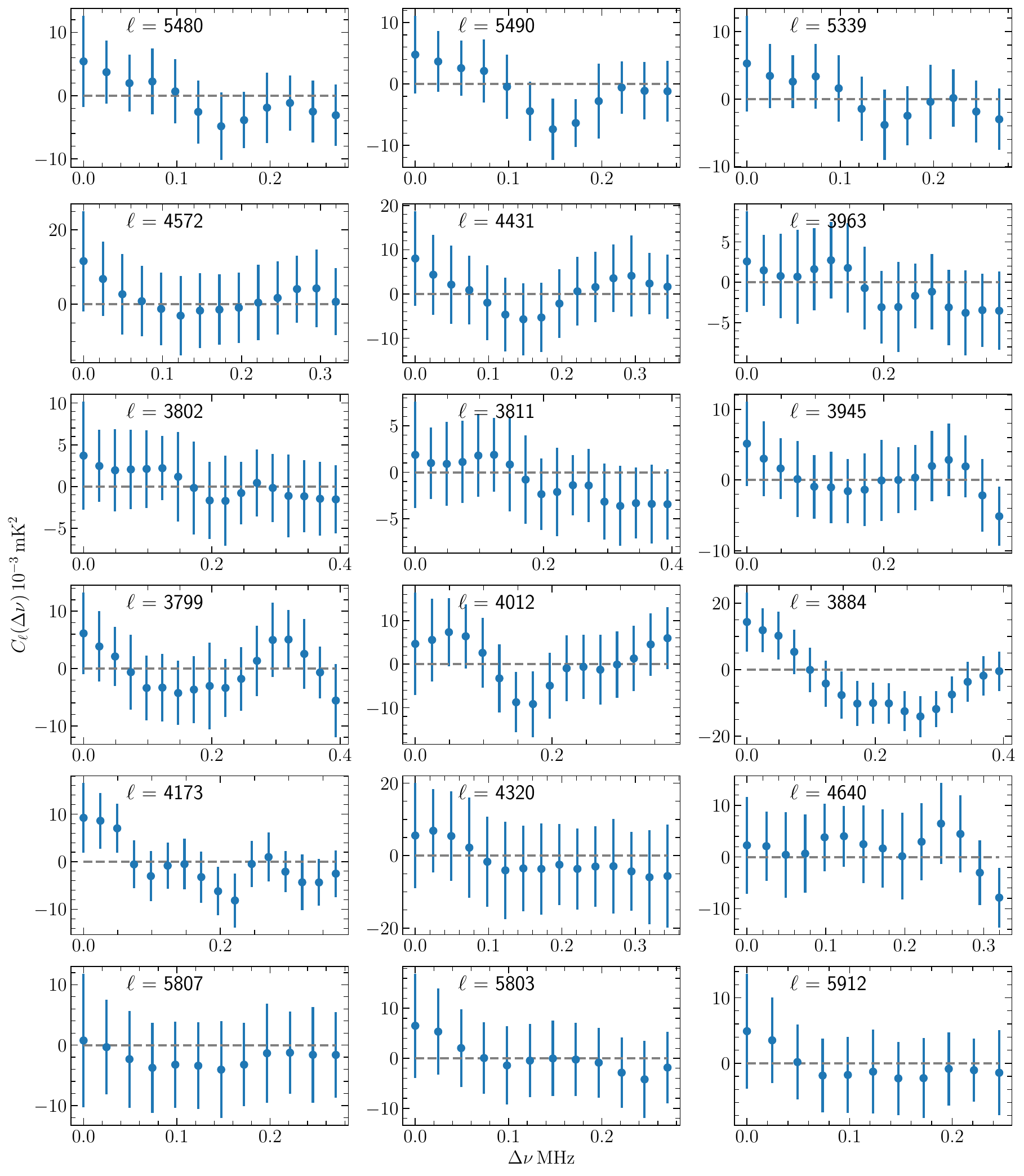}
    \caption{Same as Figure~\ref{fig:model_s1a}  but for Set III considering  the range $\Delta\nu \le  [\Delta\nu]_{0.1}$.}
    \label{fig:model_s3}
\end{figure}

\begin{figure}
    \centering
    \includegraphics[width=\columnwidth]{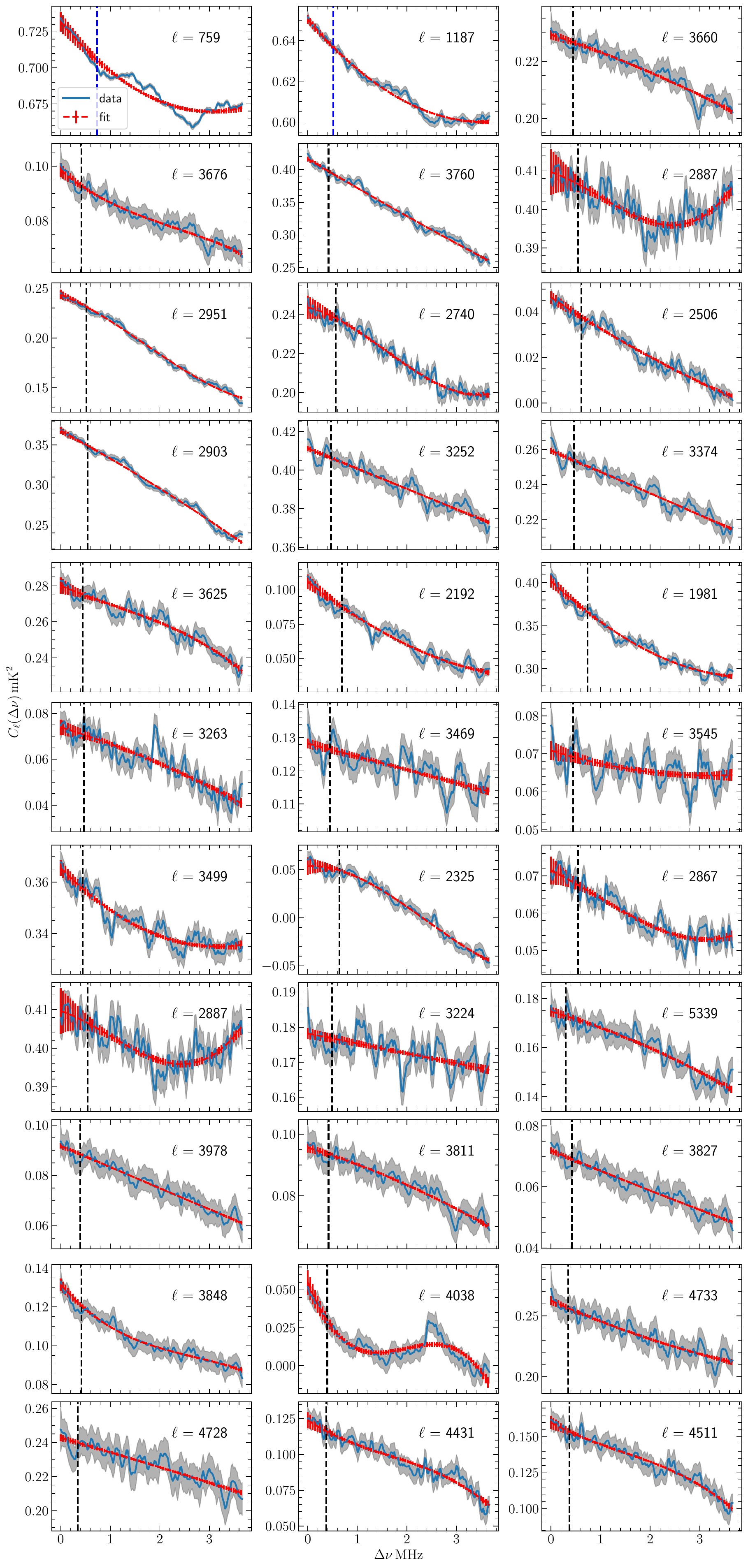}
    \caption{Same as Figure~\ref{fig:fg_fit_s1a} but for GPR and for all the sets. Here the vertical blue dashed lines show $[\Delta\nu]_{0.4}$ in the first two panels which correspond to $\ellb = 759$ and $1187$, and the vertical black dashed lines show$[\Delta\nu]_{0.1}$ in all the other panels.}
    \label{fig:fg_fit_GPR}
\end{figure}

\begin{figure}
    \centering
    \includegraphics[width=\columnwidth]{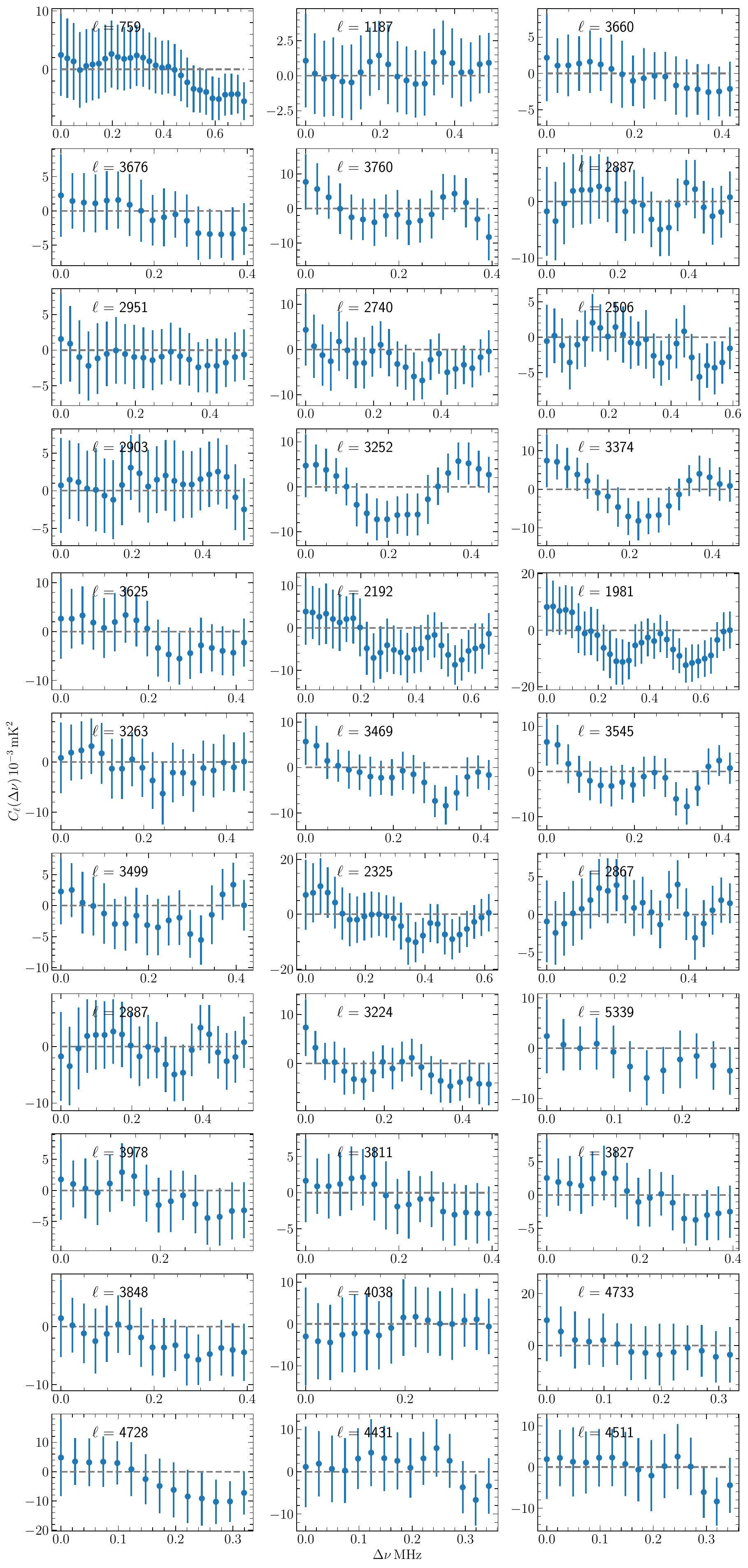}
    \caption{Same as Figure~\ref{fig:model_s1a}  but for GPR (Figure~\ref{fig:fg_fit_GPR}).}
    \label{fig:model_GPR}
\end{figure}

This appendix contains figures showing  the measured \clb{} along with  the respective  foreground models $\left[C_{\ellb}(\Delta\nu)\right]_{\rm{FG}}$ for all the unflagged $\ellb$ grid points which were used to constrain the $21$-cm signal. The corresponding residuals $\left[C_{\ellb}(\Delta\nu)\right]_{\rm{res}}$ after foreground subtraction are shown in separate figures. As mentioned in Section~\ref{sec:fitandres}, we have divided the measured \clb{} into three sets, $\ell<2000$ (Set I), $2000<\ell<4000$ (Set II) and $\ell>4000$ (Set III), and  analysed these separately. Considering PF, Figures~(\ref{fig:fg_fit_s1a}, \ref{fig:model_s1a}), (\ref{fig:fg_fit_s2}, \ref{fig:model_s2}) and (\ref{fig:fg_fit_s3}, \ref{fig:model_s3}) show the results for Sets I, II and III respectively. The results for GPR from all three sets are shown together in Figures~(\ref{fig:fg_fit_GPR}, \ref{fig:model_GPR}).




\bsp	
\label{lastpage}
\end{document}